\documentclass[]{jfm}

\usepackage{graphicx}
\usepackage{newtxtext}
\usepackage{newtxmath}
\usepackage{natbib}
\usepackage{multirow}
\usepackage{adjustbox}
\usepackage{algorithm}
\usepackage{algpseudocode}

\usepackage{hyperref}
\hypersetup{
    colorlinks = false,
    urlcolor   = cyan,
    citecolor  = cyan,
}

\newcommand{\RomanNumeralCaps}[1]

\shorttitle{Self-consistent model for active control of wind turbine wakes}
\shortauthor{Zhaobin Li and Xiaolei Yang}

\title{Self-consistent model for active control of wind turbine wakes}

\author{Zhaobin Li\aff{1,2},   \and Xiaolei Yang\aff{1,2}\corresp{\email{xyang@imech.ac.cn}}}

\affiliation{\aff{1}The State Key Laboratory of Nonlinear Mechanics, Institute of Mechanics, Chinese Academy of Sciences, Beijing 100190, China
\aff{2}School of Engineering Sciences, University of Chinese Academy of Sciences, Beijing 100049, China}

\begin{document}
\maketitle

\begin{abstract} Active wake control (AWC) has emerged as a promising strategy for enhancing wind turbine wake recovery, but accurately modelling its underlying fluid mechanisms remains challenging. This study presents a computationally efficient wake model that provides end-to-end prediction capability from rotor actuation to wake recovery enhancement by capturing the coupled dynamics of wake meandering and meanflow modification, requiring only two inputs: a reference wake without control and a user-defined AWC strategy.  The model combines physics-based resolvent modelling for large-scale coherent structures and an eddy viscosity modelling for small-scale turbulence.  A Reynolds stress model is introduced to account for the influence of both coherent and incoherent wake fluctuations, so that the time-averaged wake recovery enhanced by the AWC can be quantitatively predicted. Validation against large-eddy simulations (LES) across various AWC approaches and actuating frequencies demonstrates the model's predictive capability, accurately capturing AWC-specific and frequency-dependent mean wake recovery with less than 8\% error from LES while reducing computational time from thousands of CPU hours to minutes. The efficiency and accuracy of the model makes it a promising tool for practical AWC design and optimization of large-scale wind farms.

\end{abstract}

\begin{keywords}
Authors should not enter keywords on the manuscript, as these must be chosen by the author during the online submission process and will then be added during the typesetting process (see \href{https://www.cambridge.org/core/journals/journal-of-fluid-mechanics/information/list-of-keywords}{Keyword PDF} for the full list).  Other classifications will be added at the same time.
\end{keywords}

\section{Introduction}  

Wind farms have emerged as the predominant form of large-scale wind energy development. Within these farms, the proximity of turbines leads to significant aerodynamic interactions, particularly in the form of wake effects \citep{vermeer2003wind,stevens2017flow,Meneveau_2024}. The wake generated by upstream turbines substantially influences the inflow conditions for downstream turbines, manifesting as reduced wind speeds and increased turbulence intensity \citep{porte2020wind}, reducing both power generation and structural lifetime of wind turbines. Consequently, understanding and controlling wind turbine wakes are crucial for optimizing the economics and safety of wind farms \citep{shapiro2022turbulence} and lays the foundation for even higher levels of the electricity grid and market management \citep{eguinoa2021wind}.  Fluid mechanics is deeply involved in this research area \citep{kheirabadi2019quantitative,abkar2023reinforcement}. 

Substantial advancements in wake control have been achieved, and the research evolves from solely static wake steering to including more sophisticated active wake control (AWC) methods \citep{meyers2022wind}. While static wake steering (mainly through yaw misalignment) has demonstrated effectiveness in redirecting wakes away from downstream turbines \citep{bastankhah2016experimental,shapiro2018modelling,howland2022collective,Meneveau_2024}, AWC has emerged as a rather new and promising approach under active development. 

\subsection{State-of-the-art for AWC modelling}
AWC enhances wake recovery through well-designed dynamic perturbations of rotor force or position to generate wake pulsing or wake meandering fluctuations \citep{houck2022review}. Pulsing-type AWCs induce periodic expansion and contraction of the wake as it travels downstream. This wake pattern can be triggered through the dynamic induction control that generates oscillatory thrust forces via collective blade pitch or generator torque adjustments \citep{goit2015optimal,munters2017optimal}, or through the surge motion of floating offshore wind turbines (FOWTs) \citep{messmer2024enhanced, wang2024aerodynamic}. In contrast, meandering-type AWCs create large-scale and oscillatory lateral or helical wake deflections. Exemplary approaches are the dynamic individual pitch control (DIPC) method that triggers helical wake meandering through phase-shifted periodic variation of blade pitch angle \citep{frederik2020helix}, dynamic yaw or sway method  \citep{munters2018dynamic,lin2024wake,li2022onset} that induce horizontal wake meandering through periodic variation of yaw angle or horizontal wake position. Due to stronger cross-flow mixing and wake recovery enhancement, the meandering-type AWCs are more efficient than the pulsing-type ones  \citep{messmer2024enhanced}. For this purpose, the present work focuses on meandering-type AWCs and puts emphasis on their fast predictive modelling, which is indispensable for AWC design and optimization. To this end, recent studies on AWC modelling have developed the predictive capability mainly in predicting the optimal frequencies and fluctuation of the wake, but the predictive capability for the final target, i.e., the AWC-induced wake recovery, is still lacking. 

\begin{itemize}
    \item \textbf{Optimal frequency:} 
 Wind turbine wakes are convectively unstable and act as amplifiers of upstream perturbations, if the perturbation frequency falls within the wake's unstable frequency range   \citep{iungo2013linear,mao2018far,gupta2019low}. For this reason, the unstable frequency range of the wake can serve as a good indicator for designing optimal forcing frequency. In this regard, linear stability analysis (LSA) is often applied to the time-averaged wake field without AWC \citep{iungo2013linear,gupta2019low}.  Axisymmetric baseflow assumption is often applied in these analyses, such that the meandering-type fluctuation can be modelled by perturbations with azimuthal wave number $ \mathopen|m\mathclose|  =  1$. The unstable frequency range of wind turbine wakes falls typically within $0.2<St<0.6$  ($St = fD/U_\infty$ is the Strouhal number, with $f$ the control frequency, $D$ the rotor diameter, and $U_\infty$ the oncoming wind speed) \citep{cheung2024fluid}. Numerical simulations and experiments have demonstrated that both strong helical meandering triggered by DIPC \citep{frederik2020helix,van2024maximizing} and large side-to-side meandering triggered by rotor sway and yaw fall in this frequency range \citep{li2022onset,lin2022large,messmer2024enhanced}. Despite its success, the eigenvalue nature of the LSA \citep{Schmid2001Stability} limits its predictive capability: neither AWC-specific wake fluctuation nor enhanced wake recovery enhancement can be predicted by LSA.

 \item \textbf{Coherent velocity fluctuation:} AWC-induced wake fluctuation is characterized by large-scale coherent structures \citep{frederik2020helix,li2022onset}.  Unlike the sensitive frequency range, the coherent fluctuation is not only dependent on the wake's instability property but also on specific rotor actuation \citep{fontanella2021unaflow,kopperstad2020aerodynamic,fu2019wake}.  For this reason, wake models for predicting coherent structures must function as an input-output system that links the rotor actuation and the wake fluctuating response. The resolvent analysis \citep{trefethen1993hydrodynamic,sipp2010dynamics,mckeon2010critical} serves as an effective tool for this purpose. The resolvent analysis has been applied to explain the origin of coherent structures in wind turbine wakes generated by ambient turbulence \citep{de2022stability,feng2022componentwise,feng2024improved}. In the context of predicting rotor-forced wake response,  \cite{li2024resolvent} introduced Forcing-to-Wake model to predict coherent wake fluctuation controlled by arbitrary oscillatory rotor actuation. Validated against LES, the model successfully predicts distinct wake fluctuating structures for various FOWT-motion-induced actuations, complementing the LSA for the predictive capability of wake fluctuating response in a rotor-actuation-aware manner.  However, due to the linear nature of the resolvent operator, the effects of coherent structures on the wake recovery are  not considered by these resolvent based approaches. 

 \item \textbf{Enhanced wake recovery:} The enhanced wake recovery is the key practical benefit of AWC \citep{meyers2022wind,houck2022review}, as the wind power available in the wake is proportional to the cube of wind velocity. Recent studies have demonstrated significant wake recovery improvements: helical meandering induced by DPIC increases time-averaged wind speed, leading to a power gain of up to 7.5\% for a two-turbine wind farm \citep{frederik2020helix}, while side-to-side meandering induced by periodic sway motion accelerates wind speed recovery by up to 20\% without ambient turbulence \citep{li2022onset,messmer2024enhanced}. Based on studies of perturbed free shear flow \citep{ho1984perturbed}, these improvements occur because AWC-induced large-scale fluctuations modify the meanflow by enhancing momentum exchange between the wake and freestream. Currently, the enhanced wake recovery due to specific AWC is often evaluated using high-fidelity simulations or experiments. To the best of the authors' knowledge, no existing fast-running wake models have formulated the AWC-induced wake response as a rotor-actuation-aware input-output system for consistently predicting both coherent flow structures and the enhanced wake recovery subject to specific AWC techniques in a quantitatively accurate manner.

\end{itemize}

{\color{black}

\subsection{Resolvent analysis for predicting coherent structures and meanflow modification} \label{sec:resolvent}   

The study of AWC requires predictive models that can accurately resolve both coherent flow structures and their nonlinear interactions with the meanflow for predicting enhanced wake recovery. Recent studies have demonstrated that resolvent analysis is a viable approach for predicting coherent structures and meanflow modification of turbulent flows. This method is briefly reviewed here. 

In the fluid mechanics research community, resolvent analysis has gained significant traction for elucidating the linear amplification mechanisms that underpin the generation of coherent structures based on the meanflow. By analysing the mathematical properties of the resolvent operator derived from the linearized Navier-Stokes equations \citep{trefethen1993hydrodynamic}, this methodology quantifies how optimal harmonic forcing is amplified by the meanflow, resulting in oscillatory flow structures. The extension of this framework to turbulent meanflow \citep{mckeon2010critical} has further demonstrated that dominant large-scale coherent structures in turbulent flow can be well approximated by the leading resolvent modes, interpreted as the linear response of the meanflow to random forcing generated by the nonlinear terms in the Navier-Stokes equations. Recent advancements, such as the inclusion of eddy viscosity \citep{pickering2021optimal} and the application of sweeping-enhanced forcing \citep{wu2023composition}, improve the prediction accuracy of the flow's fluctuating response. 

In addition to predicting fluctuating responses, recent developments in stability and resolvent analysis have incorporated meanflow modifications by accounting for nonlinear interactions between coherent structures and the meanflow. \citet{mantivc2014self} explored the coupled dynamics of perturbations and meanflow in the vortex shedding phenomenon in cylinder wakes for Reynolds numbers \(Re \leq 110\). Their work computed the vortex shedding dynamics of a baseline meanflow by solving the eigenvalue problem of the linearized Navier-Stokes equations, incorporating a meanflow correction step to quantify the effect of fluctuating velocities on the meanflow. A self-consistent state is obtained when the amplitude of the fluctuating velocity renders the meanflow neutrally stable. In the input-output framework, \citet{Yim2019} proposed a self-consistent triple decomposition to predict the coherent structures and their modification effect on the meanflow forced by externally applied finite amplitude harmonic forcing derived based on the optimal forcing of resolvent analysis. This model further employs a RANS-based, meanflow-consistent eddy viscosity within both the resolvent operator and meanflow corrections to account for the effect of small-scale incoherent fluctuations. Its predictions for both coherent structures and meanflow are validated against fully nonlinear two-dimensional URANS simulations, with the same external forcing applied at Reynolds number $Re = 5 \times 10^4$. 

Besides the predictive capability, the computational efficiency  is also an important concern for developing resolvent-based models. In general, one-dimensional resolvent analyses are very efficient. The resolvent operator can be explicitly constructed by inverting the linearized Navier-Stokes equation operator, and then analysed with SVD to derive optimal forcing and response \citep{mckeon2010critical}. However, for two-dimensional problems, the explicit construction of the bi-global resolvent operator can already be limited by the prohibitive memory costs, leading to the adoption of matrix-free approaches, such as the Arnoldi algorithm \citep{martini2021efficient, kaplan2021nozzle}. Three-dimensional problems are challenging. Although advanced techniques like randomized resolvent  \citep{Ribeiro2020} and one-way Navier-Stokes methods \citep{towne2022efficient} have been proposed recently, tri-global resolvent analysis remains computationally expensive, especially when employing it in fast wind turbine wake models. For a detailed explanation of the resolvent analysis approach, readers are referred to a recent review by \citet{rolandi2024invitation}.

}
\subsection{A novel self-consistent wake model for AWC}

\begin{figure}
    \centering
    \includegraphics[width=\linewidth]{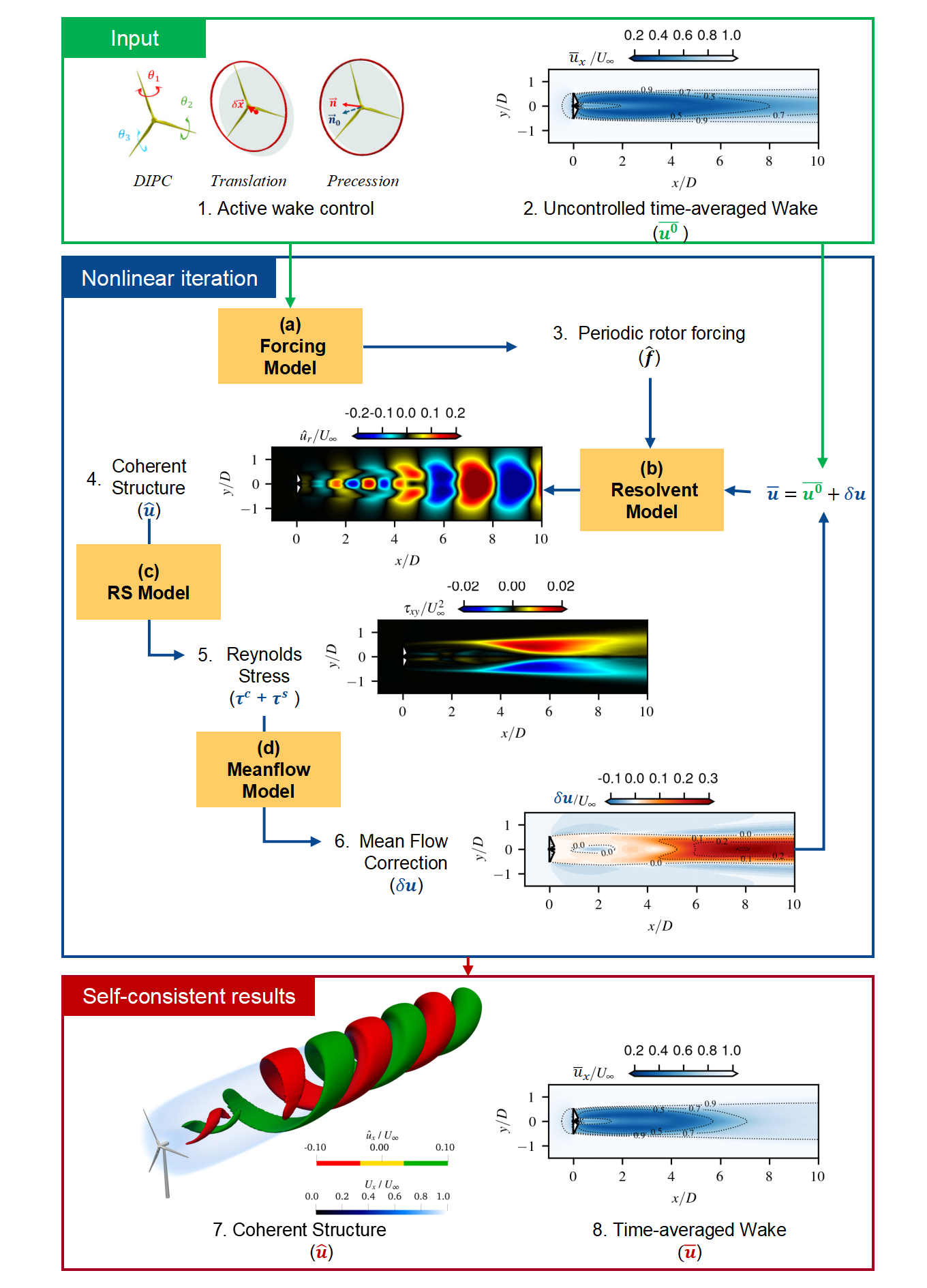}
    \caption{The proposed self-consistent wake model for AWC. See $\S$ \ref{sect:modellingFramework} for detailed explanation.}
        \label{fig:modellingFramework}
\end{figure}

The aim of this work is to develop a fast-running wind turbine wake model for self-consistent prediction of wake recovery and coherent structures, as illustrated in figure \ref{fig:modellingFramework}. The proposed model requires two inputs: the time-averaged wake of a non-controlled turbine and a specific AWC control strategy to be evaluated. The main body of the model integrates (a) a rotor forcing model for various AWC strategies, (b) a linear resolvent wake model for predicting coherent wake structures, (c) a Reynolds stress (RS) model  for coupling the fluctuation and the mean of the wake, and (d) a mean wake model. These models are coupled and solved iteratively, so the mutual influence of coherent structures and time-averaged wake is properly taken into account in a two-way manner. Upon convergence, the model provides self-consistent results on the fluctuating and time-averaged wake response. Unlike \cite{korb2023characteristics}, who studied wake deflection, deformation, and mixing in a meandering frame of reference, this work adopts an Eulerian approach to investigate the interaction between AWC-induced velocity fluctuations, RS, and meanflow modification in a fixed frame, so that the governing equations of the flow field can be more efficiently solved by conventional numerical methods. 

\begin{figure}
    \centering
    \includegraphics[width=\textwidth]{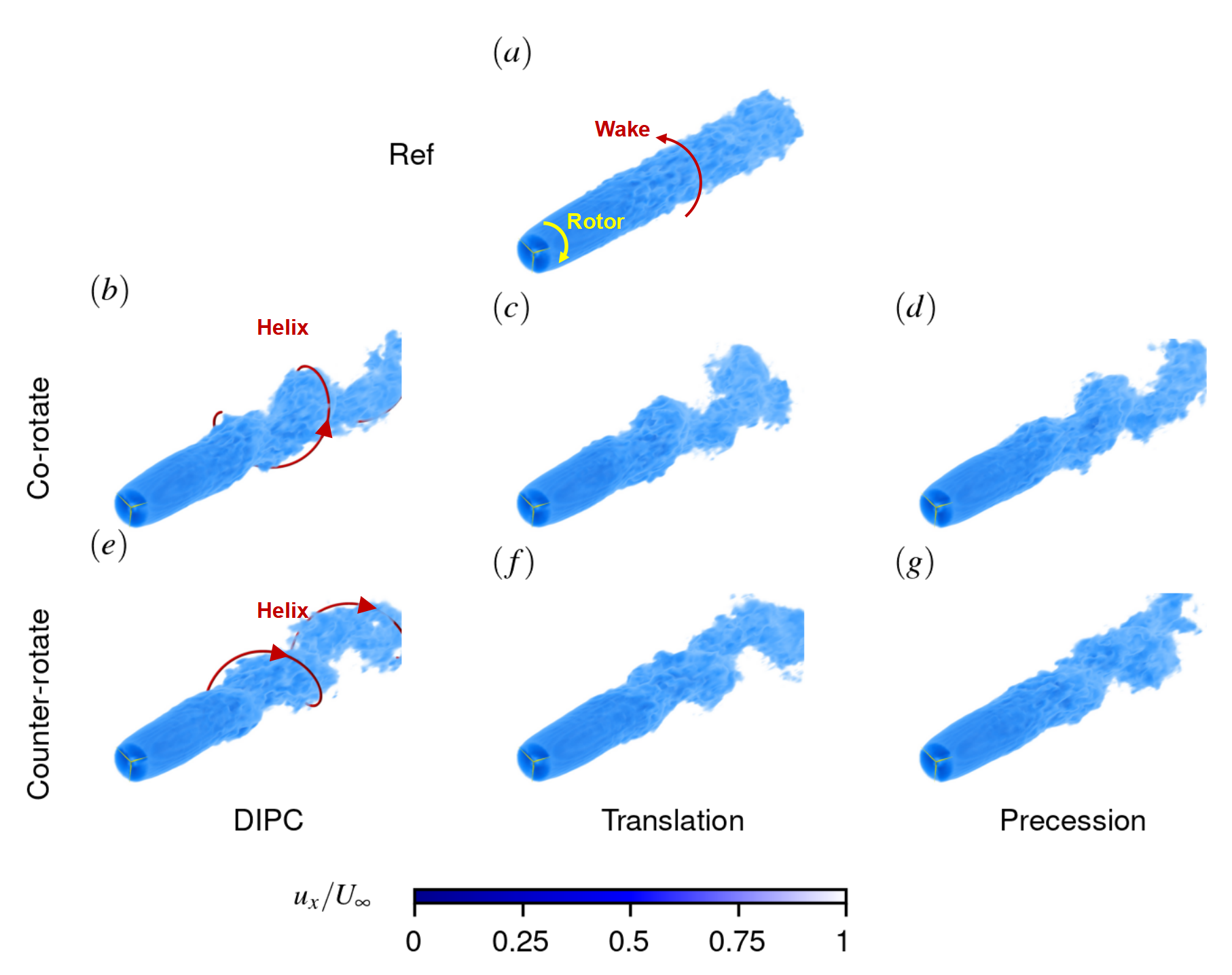}
    \caption{LES results on helical wake triggered by different AWC approaches: (a) reference uncontrolled case. (b)(e) DIPC; (c)(f) circular translation; (d)(g) rotor axis precession. The direction of the helix is defined as co-rotating when the helical structure develops in the same direction with the baseline wake while travelling downstream (b)(c)(d), and counter-rotating when opposite (e)(f)(g). Volume rendering shows streamwise velocity $u_x \approx 0.7 U_\infty$. See \S~\ref{sec:lesResult} for more details.
    \label{fig:InstantanousWake}}
    %
\end{figure}

As a first step, we focus on AWC techniques that generate helically deflected wakes, as illustrated in figure \ref{fig:InstantanousWake}. This pattern is selected for its efficiency in enhancing wake recovery \citep{frederik2020helix} and because it represents the basic fluctuating structure of an axisymmetric wake,  so that the resolvent analysis can be formulated and solved in a bi-global axisymmetric framework.  LES with rotating blades modelled as actuator surfaces is employed to verify critical modelling assumptions and to validate the proposed model across various AWC approaches and forcing frequencies.  More details about the AWC approaches, the LES method, and the result presented in the figure will be explained in the main text. 

{\color{black} This work has three primary objectives. First, a computationally efficient wake model will be developed to predict the wake response to AWC within minutes compared to conventional LES that typically demands thousands of CPU hours. Second, we will establish and validate the pure Eulerian modelling framework that represents wake meandering as large-scale velocity oscillations for investigating their effects on mean wake recovery, providing an alternative to commonly applied Lagrangian approaches to track the meandering motion of wake deficit \citep{larsen2008wake}. Third, the model focuses on the contribution of the far-wake shear layer Kelvin-Helmholtz instability actuated by rotor-level low-frequency forcing and seeks to demonstrate its dominant role in AWC-enhanced wake recovery.}

The remainder of this paper is organized as follows: \S~\ref{sec:nonlinearModel} describes the framework of the proposed wake model and details the theoretical derivation of each component. \S~\ref{sec:ModelValidation} employs a series of LES to analyse the physics of AWC-induced wake response, validate key modelling assumptions, and demonstrate the model's predictive capabilities. The computational cost, fluid physics, limitations and potential refinements of the proposed model will be discussed in \S~\ref{sec:discussion}. Finally, \S~\ref{sec:conclusion} provides concluding remarks.

\section{Self-consistent wake model \label{sec:nonlinearModel}}

\subsection{Modelling framework and assumptions \label{sect:modellingFramework}}

The proposed modelling framework is presented schematically in figure \ref{fig:modellingFramework}. The schematic distinguishes between data sets, denoted by numerical indicators (1 to 8), and mathematical models, represented by alphabetical labels (a to d). The inputs, main body of the model, and the outputs are explained as follows:

\begin{itemize}
    \item \textbf{Input:} To evaluate the wake response to a specific AWC approach, the two inputs of the model are: (1) the AWC approach to be evaluated and (2) a non-controlled baseline wake.   
    \item \textbf{Main body:} The main body of the model starts with (a) a forcing model for transferring the AWC into (3) equivalent periodic rotor forcing. Then an iterative loop starts, involving three interconnected components, including: (b) a resolvent model for predicting (4) the coherent structures in the wake ($\hat{\boldsymbol{u}}$); (c) a Reynolds stress model that predicts (5) the RS generated by both coherent structures ($\boldsymbol{\tau}^c$) and small-scale turbulence ($\boldsymbol{\tau}^s$); (d) a meanflow model to compute (6) the meanflow correction, i.e., the enhanced wake recovery ($\delta \boldsymbol{u}$), which is added back to (2) the non-controlled time-averaged wake to produce the baseflow for the resolvent analysis in the next iteration. 
    \item \textbf{Output:} Upon achieving convergence of the iteration loop, the model yields self-consistent predictions of: (7) the AWC-induced coherent structures and (8) the time-averaged wake. 
\end{itemize}

The main assumptions for the modelling include: (i) the periodicity of forcing and response, i.e., both are simple harmonic oscillations in time, required by the resolvent model; (ii) the axisymmetry of the baseflow, i.e., no shear or veer of incoming wind are considered. The former assumption facilitates the solution in frequency domain, and the latter reduces the wake to a two-dimensional problem for enhancing the model's efficiency.    

{\color{black} In the present work, an LES with a wind turbine modelled with an actuator surface method (see \S~\ref{sec:LES}) is conducted to obtain the non-controlled baseline wake and the time-averaged aerodynamic forces on the rotor.} 

\subsection{Governing equations \label{sec:mathformulation}}

In this work, we seek to predict the mean and the large-scale coherent fluctuations of the wake. To this end, we employ the triple decomposition  \citep{hussain1970mechanics, Yim2019} to separate the different components of the flow variables as follows,
\begin{align}
    \boldsymbol{u} & = \overline{\boldsymbol{u}} + \boldsymbol{u}^{c}(t) + \boldsymbol{u}^{s}(t) = \widetilde{\boldsymbol{u}}(t) + \boldsymbol{u}^{s}(t),  \\
    p & = \overline{p} +  p^{c}(t) + p^{s}(t) = \widetilde{p}(t) + p^{s}(t) , \\
    \boldsymbol{f} & = \overline{\boldsymbol{f}} +  \boldsymbol{f}^{c}(t) +  \boldsymbol{f}^{s}(t) = \widetilde{\boldsymbol{f}}(t) + \boldsymbol{f}^{s}(t). 
\end{align}
where $\boldsymbol{u}$, $p$, $\boldsymbol{f}$ represent velocity, pressure, and rotor forcing, respectively. The symbol $\overline{\{\cdot\}}$ represents the time-averaged part, $\{\cdot\}^{c}$ denotes the coherent fluctuating part, $\{\cdot\}^{s}$ denotes the small-scale turbulent part.  $\widetilde{{\{\cdot\}}}$ denotes the filtered field containing both mean and coherent fluctuation, while the small-scale turbulent component is excluded. As the AWC focuses on the mean and the large-scale coherent fluctuation of the wake, we focus on the filtered fields and assume that they verify the incompressible Navier-Stokes equation as follows,  
\begin{align}
        \frac{\partial  \widetilde{\boldsymbol{u}}}{\partial t}+ (\widetilde{\boldsymbol{u}} \cdot \nabla) \widetilde{\boldsymbol{u}}  + \widetilde{p} - \nu_\text{eff} \nabla^2 \widetilde{\boldsymbol{u}} & = \widetilde{\boldsymbol{f}}, \label{eqn:nsMom}\\
    \nabla \cdot  \widetilde{\boldsymbol{u}} &= 0, \label{eqn:nsCon}
\end{align}
Here, an effective viscosity $\nu_\text{eff}$ is introduced to account for the influence of small-scale turbulence \citep{symon2023use}. The closure will be provided in $\S$ \ref{sec:RSSModel} through an eddy viscosity model.


By assuming that both the coherent parts of forcing ($\boldsymbol{f}^c$) and response ($\boldsymbol{u}^c$, ${p}^c$) are simple harmonic, they are written as:  
\begin{equation}
[ \boldsymbol{u}^c(t), p^c(t), \boldsymbol{f}^c(t)] = \frac{1}{2}[\hat{\boldsymbol{u}}, \hat{p}, \hat{\boldsymbol{f}}]  \text{e}^{-\mathrm{i} \omega t} + \frac{1}{2} [\hat{\boldsymbol{u}}^H, \hat{p}^H, \hat{\boldsymbol{f}}^H]  \text{e}^{\mathrm{i} \omega t}   \label{eqn:spaceTimeDecomposition}
\end{equation}
where  $\mathrm{i} = \sqrt{-1}$ is the imaginary unit, $\omega$ is the angular frequency, and $\{\cdot\}^H$ denotes complex conjugate. The spatial modes are represented by $\hat{\{\cdot\}}$, while $\text{e}^{\pm \mathrm{i} \omega t}$ describes the temporal oscillation. 

Substituting this ansatz into the equations \eqref{eqn:nsCon} and separating coherent fluctuating and time-averaged components yields two coupled systems.  The dynamics of the coherent fluctuation at forcing frequency $\omega$ is described by the harmonic linearized Navier-Stokes equation:
\begin{align}
    -\mathrm{i} \omega \hat{\boldsymbol{u}} + (\overline{\boldsymbol{u}} \cdot \nabla) \hat{\boldsymbol{u}} +  (\hat{\boldsymbol{u}} \cdot \nabla) \overline{\boldsymbol{u}} + \nabla \hat{p} - \nu_\text{eff} \nabla^2 \hat{\boldsymbol{u}} & = \hat{\boldsymbol{f}}, \label{eqn:coherentStructuresMom} \\
    \nabla \cdot \hat{\boldsymbol{u}} & = 0.  \label{eqn:coherentStructuresIncom}
\end{align}
These equations govern the resolvent model  \citep{li2024resolvent}, which will be detailed in \S \ref{sec:resolventModel}.

For the time-averaged field, the governing equations are:
\begin{align} 
 \left(\overline{\boldsymbol{u}} \cdot \nabla \right) \overline{\boldsymbol{u}}  + \nabla \overline{p} & = \overline{\boldsymbol{f}} + \nabla \cdot \boldsymbol{\tau^c} +  \nu_\text{eff} \nabla^2 \overline{\boldsymbol{u}}, \label{eqn:meanMom} \\
    \nabla \cdot  \overline{\boldsymbol{u}} &= 0.  \label{eqn:meanIncom}
\end{align}
Here, $\boldsymbol{\tau^c} = \tau^c_{ij}  = -\overline{u^c_{i}u^c_{j}}$ represents the Reynolds stress tensor arising from AWC-induced coherent fluctuation (${\boldsymbol{u}^c}$), where indices $i,j \in \{1,2,3\}$ denote velocity components in three-dimensional space. Once coherent fluctuations are excited by AWC, $\nabla \cdot \boldsymbol{\tau^c}$ drives meanflow modification. $\nu_\text{eff} \nabla^2 \overline{\boldsymbol{u}}$ represents the effect of  small-scale turbulence on the mean wake. These two terms are included in the Reynolds stress model and will be explained in \S \ref{sec:meanFlowCorrModel}.

The above two equation systems are interconnected: the coherent fluctuation of the wake is computed with equations \eqref{eqn:coherentStructuresMom} and \eqref{eqn:coherentStructuresIncom} based on the mean wake $\overline{\boldsymbol{u}}$; and the mean wake is also influenced by the coherent fluctuation through  RS ($\boldsymbol{\tau}^c$) in equation \eqref{eqn:meanMom}.

\subsection{Rotor forcing model \label{sec:forcingModel}}

\begin{figure}
    \centering
    \includegraphics[width=0.8\linewidth]{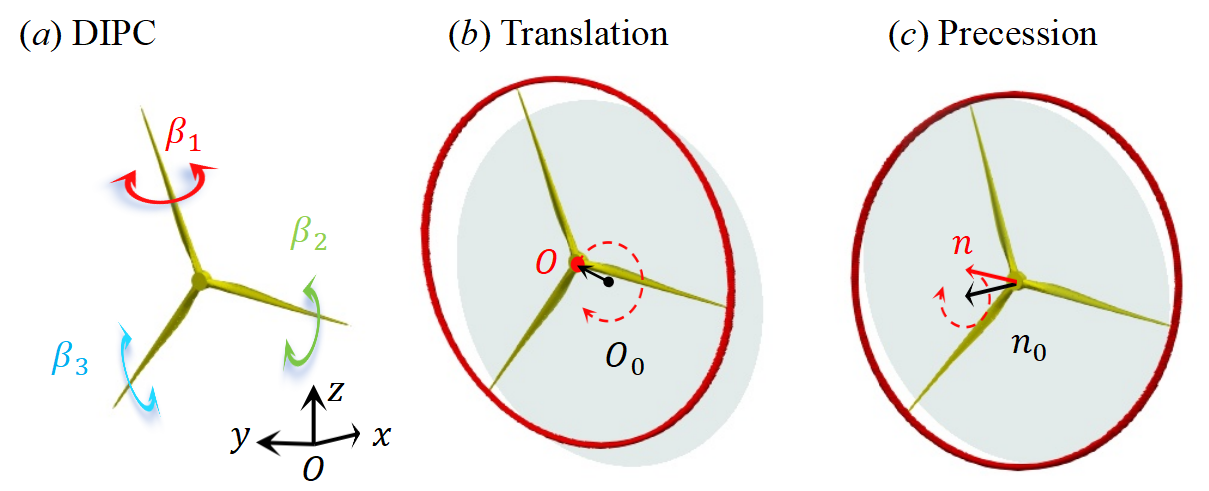}
    \caption{Schematic of AWC approaches: (a) DIPC; (b) in-plane circular translation; (c) precession of the rotor. Shaded areas in (b) and (c) indicate the original rotor swept area.}
    \label{fig:AWC_Schema_01}
\end{figure}

In the present framework, the rotor forcing model serves to transfer the different AWC approaches to coherently oscillatory forcing $\hat{\boldsymbol{f}}$, which will be employed as an input of the resolvent model when solving equation \eqref{eqn:coherentStructuresMom}. Here, we develop analytical forcing models for three different AWC approaches that induce helical-type meandering, namely the dynamic individual pitch control, in-plane circular translation, and precession of the rotor, as shown in figure  \ref{fig:AWC_Schema_01}. The first approach introduces wake perturbation through DIPC, which is proposed by \cite{frederik2020helix} and is known as the helix approach. The latter two approaches perturb the wake via rotor displacements, which exist in FOWT due to platform motion \citep{tran2015aerodynamic}.  Following the axisymmetry assumption of the time-averaged wake, we model only the global effects at the rotor disk level rather than individual blade effects, and assume small perturbation amplitudes to enable linearization of rotor aerodynamics about the base state.

\subsubsection{Dynamic individual pitch control \label{sec:DIPC}}

As shown in figure \ref{fig:AWC_Schema_01} (a), the DIPC approach periodically adjusts the pitch angle $\beta$ of each blade at a common frequency $\omega_p$, with well-designed phase differences between blades, as follows:
\begin{equation}
    \beta_k(t) = A_\beta \sin(\omega_p t + \frac{2\pi}{3}(k-1)), \label{eqn:dynamicPitch}
\end{equation}
where $k \in \{1,2,3\}$ is the blade index and $A_{\beta}$ is the angular amplitude of pitch control. The angular velocity of the pitch variation is $\omega_p = \Omega \pm \omega$ with $\Omega$ the angular velocity of the rotor. This control strategy leads to periodic variation of the angle of attack (AoA) and the aerodynamic forces. The first blade points upwards (in $z$ direction) at $t = 0$, and its azimuth angle is $\psi_1 = \Omega t$. According to the Multi-Blade Coordinate transformation \citep{bir2008multi,van2024maximizing}, {\color{black} DIPC introduces an equivalent rotor level rotating radial force \citep{korb2023characteristics}}.  A linear relationship between the force and the pitch angle variations has been proposed \citep{korb2023characteristics}, but the coefficient preceding the linear relation is undetermined.  Here we propose an approach to determine this coefficient by using simply the blade aerodynamic data obtained from the non-controlled baseline case.  

The AoA of a blade under DIPC varies as a function of time $(t)$ and radial location $(r)$:
\begin{equation}
    \alpha_k (t,r) =  \alpha^0(r) - \beta_k(t). 
\end{equation}
$\alpha^0 (r)$ is the AoA of the baseline case without AWC and $\beta_k(t)$ is the pitch variation defined in equation \eqref{eqn:dynamicPitch}. Following \cite{korb2023characteristics}, we consider only the contribution from the lift, whose variation due to  DIPC can be expressed as:
\begin{equation}
    \begin{split}
        \frac{\delta L_k(t,r)}{\overline{L^0}(r)} & =  \frac{ C_L(\alpha_k(t,r))  -   C_L(\alpha^0(r))}{C_L(\alpha^0(r))}  \\
        & \approx  - \frac{\beta_k(t)}{C_L(\alpha^0(r))}  \left.\frac{\partial C_L}{\partial \alpha}\right|_{\alpha^0(r)}. \\
    \end{split}
\end{equation}
Here, $\overline{L^0}$ denotes the baseline time-averaged lift force and $\delta L_k$ denotes the lift variation. This expression employs a linearization  around the baseline state using Taylor expansion, assuming a small pitch variation. By using the geometric similarity of force components, this relationship can be projected into the streamwise $(x)$ and azimuthal $(\theta)$ directions to obtain the thrust and tangential force variations:
\begin{equation}
    \frac{\delta F_{x,k}(t,r)}{\overline{F^0_{x}}(r)} = \frac{\delta F_{\theta,k}(t,r)}{\overline{F^0_{\theta}}(r)} = \frac{\delta L_k(t,r)}{\overline{L^0}(r)} =- \frac{\beta_k(t)}{C_L(\alpha^0(r))}  \left.\frac{\partial C_L}{\partial \alpha}\right|_{\alpha^0(r)} \label{eqn:DIPC_blade}.
\end{equation}
Once the baseline AoA ($\alpha^0(r)$) and the baseline blade forces ($\overline{F^0_x}(r)$,  $\overline{F^0_\theta}(r)$) are obtained, the force variation can be quantitatively computed with the derivative of the lift coefficient $\left(\displaystyle  \left.\frac{\partial \overline{ C_L}}{\partial \alpha}\right|_{\alpha^0(r)}\right)$ from the airfoil database and the pitch variation ($\beta_k(t)$).

Next, the aerodynamic force variation is transformed from the blade-following reference frame to a ground-fixed one, to model the effect of DIPC on a stationary actuator disk exerting equivalent unsteady normal and tangential forces. To achieve this, we assume that the rotor's rotational frequency is significantly larger than the forcing frequency ($|\Omega| \gg |\omega|$). This assumption allows us to neglect the time interval between blade passages and to average the forces on the rotor swept area. {\color{black}At} any point $(r, \theta)$ on the rotor swept area, the DIPC-controlled rotor, as described in equation \eqref{eqn:dynamicPitch}, generates a streamwise blade force at three blade passages per revolution ($t_1$, $t_2$, $t_3$) as follows: 
\begin{equation}
    \delta F_{x,k}(t_k,r)  = - \frac{A_\beta \overline{F^0_x}(r)}{C_L(\alpha_0(r))}  \left.\frac{\partial C_L}{\partial \alpha}\right|_{\alpha_0(r)} \sin{ \left( \left(\Omega \pm \omega \right) t_k  +  \frac{2\pi}{3}(k-1)  \right) } 
\end{equation}
with $k \in {1,2,3}$ the blade index. The blade passage instant is expressed as:
\begin{equation}
    t_k = \frac{\theta - \frac{\pi}{2}- \frac{2\pi}{3}(k-1) }{\Omega}.
\end{equation}
Combining these equations, yields, 
\begin{equation}
     \delta F_{x,k}(t_k,r) = \frac{A_\beta \overline{F^0_x}(r)}{C_L(\alpha_0(r))}  \left.\frac{\partial C_L}{\partial \alpha}\right|_{\alpha_0(r)} \cos{ \left(\theta \pm \omega t_k \right)} 
\end{equation}
By assuming $t_k \approx t$ and averaging for a period of rotor evolution, the streamwise coherent force variation per unit area exerted by the equivalent actuator disk on the flow can be expressed as: 
\begin{equation}
f^c_x  = -A_\beta \cos(\theta \pm \omega t) \frac{1}{2} U_\infty^2 \frac{\overline{C^0_x}(r)}{C_L(\alpha_0(r))} \left.\frac{\partial C_L}{\partial \alpha}\right|_{\alpha_0(r)}, 
\end{equation}
where {\color{black} ${\overline{C^0_x}(r)}$ denotes the local thrust coefficient at radial position $r$}, defined as:
\begin{equation}
        \overline{C^0_x}(r) = \frac{3 \cdot \overline{F^0_x}(r) \text{d} r}{ 2\pi r \text{d} r \cdot \frac{1}{2} U_\infty^2} =  \frac{3 \overline{F^0_x}(r) }{ \pi r   U_\infty^2 }. 
    \label{eqn:ft_define}
\end{equation}
Note that the sign of the force on the blade and that on the flow are opposite.

Similarly, the coherent tangential force variation per unit area exerted by the actuator disk can be calculated as:
\begin{equation}
f^c_\theta  = -A_\beta \cos(\theta \pm \omega t) \frac{1}{2} U_\infty^2 \frac{\overline{C^0_\theta}(r)}{C_L(\alpha_0(r))} \left.\frac{\partial C_L}{\partial \alpha}\right|_{\alpha_0(r)}, 
\end{equation}
with $\overline{C^0_\theta} (r)$ the time-averaged tangential force coefficient of the baseline case. By assuming radial force variation is negligible, the rotor force variation can be expressed as: 

\begin{equation}
    \boldsymbol{f^c}(t) =  -\frac{A_\beta \cos(\theta \pm \omega t) U_\infty^2}{2 C_L(\alpha_0(r))} \left.\frac{\partial C_L}{\partial \alpha}\right|_{\alpha_0(r)} \left[\overline{C^0_x}(r), 0, \overline{C^0_\theta}(r) \right]
\end{equation}

\begin{figure}
    \centering
    \includegraphics[width=0.75\linewidth]{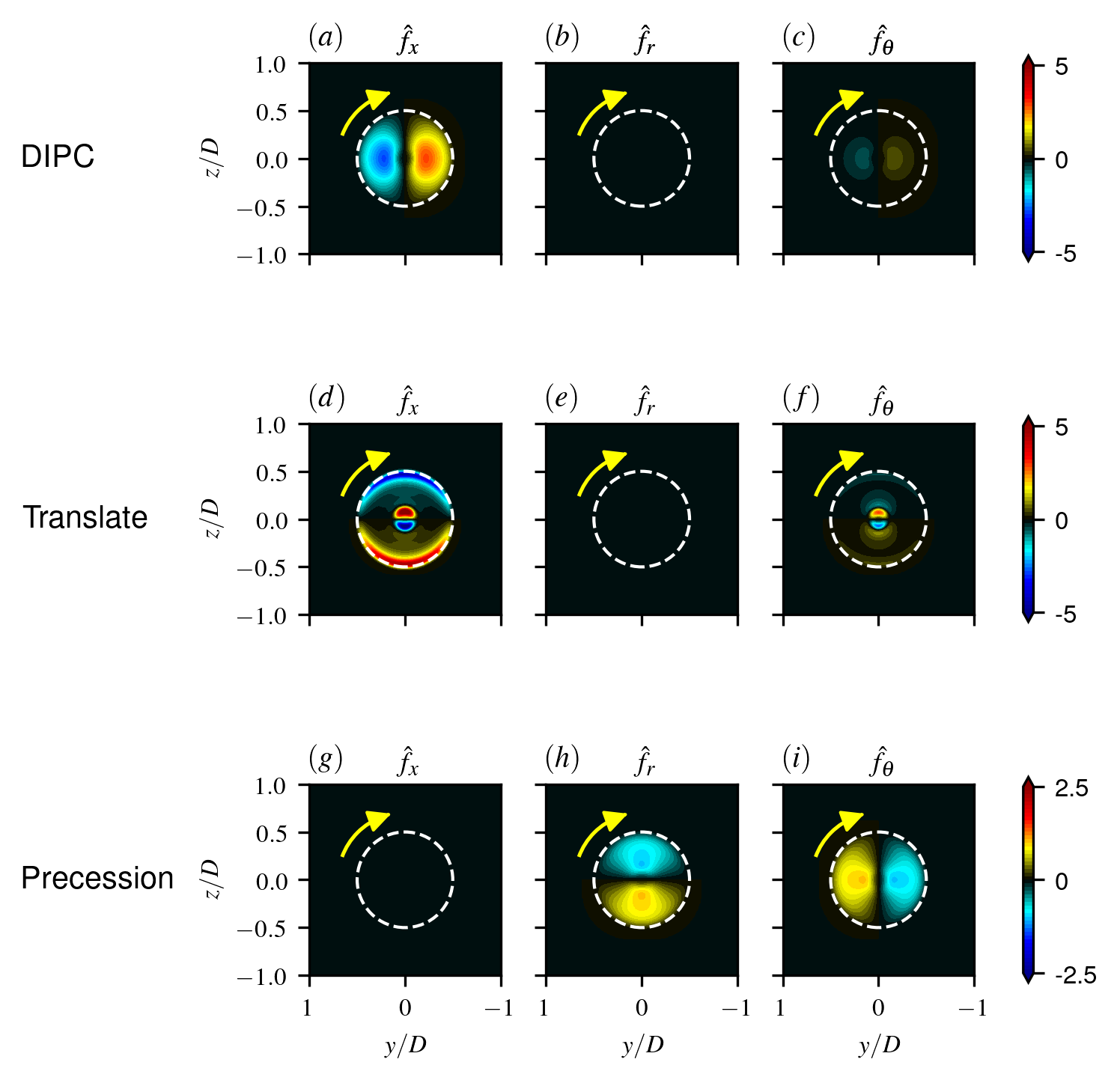}
    \caption{The spatial modes of unsteady rotor forcing on the actuator disk for DIPC (a–c), translation (d–f), and precession (g–i). The left, centre, and right columns display the streamwise, radial, and azimuthal components, respectively. The forcings are normalized by $A U^2_\infty$, where $A$ is the amplitude of each control strategy. Yellow arrows indicate the rotating direction of the rotor. }   
    \label{fig:RotorForcingModes}
\end{figure}

The frequency-domain transformation of these forces reveals their spatial distribution, as shown in figure \ref{fig:RotorForcingModes} (a–c). The streamwise component dominates and exhibits anti-symmetry about the rotor diameter, generating an out-of-plane moment in the wake flow \citep{van2024maximizing}. It rotates around the rotor's axis over time, providing initial perturbation to the wake.

\subsubsection{{Circular translation}}

 In-plane circular translation involves circular movement of the rotor within its rotation plane (figure \ref{fig:AWC_Schema_01} b). The hub centre traces a circular path of radius $A_t$ in the $yz$ plane relative to its original position ($O_0$). This motion is equivalent to a combination of synchronized sway and heave movements, with horizontal and vertical displacements expressed as:
    \begin{align}
    \delta y (t) &= A_{t}\sin(\omega t), \label{eqn:ct1} \\
    \delta z (t) &= \pm A_{t}\cos(\omega t). \label{eqn:ct2}
    \end{align}
Here, the sign of the vertical displacement $\delta z (t)$ will define the helix direction of the wake.  The associated unsteady rotor forcing can be determined straightforwardly by using the motion-to-forcing model proposed in \citep{li2024resolvent}, which computes the aerodynamic force variation due to harmonic rotor position oscillations. The resulting thrust and tangential force oscillations are: 
\begin{equation}
         f^c_x  = \frac{U_\infty^2 }{2} \frac{\partial {\overline{C^0_x}}(r)}{\partial r} \left( \delta y(t) \cos{\theta}  +  \delta z(t) \sin{\theta}  \right) 
         =  A_t \frac{U_\infty^2}{2}  \frac{\partial {\overline{C^0_x}}(r)}{\partial r} \sin (\pm \theta + \omega t)
\end{equation}
and 
\begin{equation}
     f^c_\theta = \frac{U_\infty^2 }{2} \frac{\partial {\overline{C^0_\theta}}(r)}{\partial r} \left( \delta y(t) \cos{\theta}  +  \delta z(t) \sin{\theta}  \right)  = A_t  \frac{U_\infty^2}{2}  \frac{\partial {\overline{C^0_\theta}}(r)}{\partial r} \sin (\pm \theta + \omega t)
\end{equation}
The sign $\pm$ preceding $\theta$ determines the azimuthal direction of the circular in-plane translation. The radial forcing is neglected. 

The spatial modes of the forcing are shown in figure \ref{fig:RotorForcingModes} (d-f). As seen, the motion generates an antisymmetric streamwise forcing concentrated near the rotor's edge and centre, corresponding to the location where the radial gradient of the thrust is highest. The tangential forcing appears mainly near the wake centre.

\subsubsection{Rotor precession}

Precession refers to the circular motion of the turbine rotor's axis, which traces out a cone-like path, as shown in figure \ref{fig:AWC_Schema_01} (c). This movement results in a misalignment of the rotor axis out of the wind direction, generating a side force to the wake. The normal direction of the rotor is defined as 
        \begin{align}
        n_x & = -\sqrt{1 - A_\gamma^2},    \label{eqn:cr1} \\
        n_y(t) & = A_\gamma \sin(\omega t), \label{eqn:cr2}\\
        n_z(t) & = \pm A_\gamma \cos(\omega t),  \label{eqn:cr3}
    \end{align}
where $A_\gamma$ is the misalignment between the rotor axis and the oncoming wind. The above expression indicates that the precession can be considered as a combined horizontal and vertical side-to-side oscillatory rotor inclination.  The precession direction and the wake helix direction are both determined by the sign of $n_z (t)$ in equation \eqref{eqn:cr3}. 

Analogously to the circular translation, the forcing generated by the circular rotation can also be obtained straight-forwardly by the motion-to-forcing model \citep{li2024resolvent} as combine yaw and tilt motions. By incorporating equations \eqref{eqn:cr1} to \eqref{eqn:cr3}  into the motion-to-forcing model, the radial and tangential components of the rotor forcing are expressed as follows:
\begin{align}
    f^c_r & = -\frac{A_\gamma}{2} U_\infty^2 {\overline{C^0_x}}(r) \sin\left( \pm \theta - \omega t \right), \\
    f^c_\theta & = -\frac{A_\gamma}{2} U_\infty^2 {\overline{C^0_x}}(r) \cos\left( \pm \theta - \omega t \right),
\end{align}
Here, $A_\gamma$ represents the angle of the rotor's misalignment. The sign preceding $\theta$ is dependent on the movement azimuthal direction. For this movement, the streamwise forcing is one order smaller than the radial and tangential components \citep{li2024resolvent}, and thus it is neglected here. The spatial modes are shown in figure \ref{fig:RotorForcingModes} (g-i).

\subsection{Resolvent model\label{sec:resolventModel}}

The resolvent model is employed to predict the coherent wake fluctuation induced by AWC. As shown in figure \ref{fig:modellingFramework}, the model requires two inputs: (i) a rotor forcing representing the AWC-induced perturbation, computed by the rotor-forcing model $\S$ \ref{sec:forcingModel} and (ii) a time-averaged baseflow $\overline {\boldsymbol{u}}$. The model outputs the spatial modes of coherent wake fluctuation ($\hat{\boldsymbol{u}}$). While the inputs and outputs match the previously proposed forcing-to-wake model exactly, our implementation requires updating the resolvent operator in each iteration using the modified time-averaged flow field. In the following, the key principles of the resolvent model will be presented. For a detailed understanding of the fundamental principles, readers may refer to \cite{li2024resolvent} for the forcing-to-wake model and \cite{rolandi2024invitation} for broader context on resolvent analysis.


The resolvent operator is derived from the harmonic linearized Navier-Stokes equations that govern the harmonic fluctuating part of the flow, i.e., equations \eqref{eqn:coherentStructuresMom} and \eqref{eqn:coherentStructuresIncom}. Writing these equations in matrix form, yields: 
\begin{equation}
\left( -i\omega \begin{bmatrix} \boldsymbol{I} & \\ & 0  \end{bmatrix} - \boldsymbol{L}(\overline{\boldsymbol{u}}) \right)
\begin{bmatrix}
  \hat{\boldsymbol{u}}  \\
  \hat{p}
\end{bmatrix}
     =    \begin{bmatrix}
  \hat{\boldsymbol{f}}  \\
0
\end{bmatrix}. \label{eqn:resolventEquation}
\end{equation}
Here, $\boldsymbol{I}$ represents a $3 \times 3$ identity matrix arising from the temporal derivatives of $\boldsymbol{u}^c$, while  $\boldsymbol{L}(\overline{\boldsymbol{u}})$ encompasses all remaining linear operations acting on the spatial modes of the fluctuating velocity $\hat{\boldsymbol{u}}$ and pressure $\hat{p}$ in equations \eqref{eqn:coherentStructuresMom} and \eqref{eqn:coherentStructuresIncom}. The resolvent operator $\boldsymbol{R}(\overline{\boldsymbol{u}})$ is expressed as: 
\begin{equation}
    \boldsymbol{R}(\overline{\boldsymbol{u}}) = \left( -i\omega \begin{bmatrix} \boldsymbol{I} & \\ & 0  \end{bmatrix} - \boldsymbol{L}(\overline{\boldsymbol{u}}) \right)^{-1}.  
\end{equation}
Using this operator, the coherent fluctuations $[\hat{\boldsymbol{u}}, \hat{p}]$ are directly linked to the AWC-induced rotor forcing $\hat{\boldsymbol{f}}$, as follows:
\begin{equation}
\begin{bmatrix}
  \hat{\boldsymbol{u}}  \\
  \hat{p}
\end{bmatrix}
     =  \boldsymbol{R}(\overline{\boldsymbol{u}})  \begin{bmatrix}
  \hat{\boldsymbol{f}}  \\
0
\end{bmatrix}. \label{eqn:resolventEquation2}
\end{equation}
To improve computational efficiency, we assume an axisymmetric mean wake and apply Fourier transform in the azimuthal direction, reducing the problem to the two inhomogeneous directions $(x,r)$ for spatial  discretization: 
\begin{equation}
\begin{bmatrix}
\boldsymbol{u}^c (x,r,\theta,t) \\  p^c (x,r,\theta,t) \\ \boldsymbol{f}^c (x,r,\theta,t)
\end{bmatrix}   = \frac{1}{2}
\begin{bmatrix} 
\hat{\boldsymbol{u}}(x,r) \\ \hat{p}(x,r) \\ \hat{\boldsymbol{f}}(x,r)
\end{bmatrix} \text{e}^{\mathrm{i} (m \theta -\omega t)} + \frac{1}{2} \begin{bmatrix}
    \hat{\boldsymbol{u}}^H(x,r) \\ \hat{p}^H(x,r) \\ \hat{\boldsymbol{f}}^H(x,r)
\end{bmatrix}  \text{e}^{-\mathrm{i} (m \theta - \omega t)},  \label{eqn:azimuthalAntaz}  
\end{equation}
where $m$ is the azimuthal wave number. Substituting this ansatz into the linearized Navier-Stokes equations \eqref{eqn:coherentStructuresMom} and \eqref{eqn:coherentStructuresIncom} yields: 
\begin{align}
-\textrm{i}\omega \hat{u}_x  & =  -\Omega \hat{u}_x  -\hat{u}_r \frac{\partial \overline{u}_x}{\partial r}-\hat{u}_x \frac{\partial \overline{u}_x}{\partial x} -  \frac{\partial \hat{p}}{\partial x}+ \nu_\text{eff} (\Delta + \frac{1}{r^2}) \hat{u}_x +\hat{f}_x, \label{eqn:LNS1}\\
 -\textrm{i}\omega \hat{u}_r & = -\Omega \hat{u}_r  -\hat{u}_r \frac{\partial \overline{u}_r}{\partial r}-\hat{u}_x \frac{\partial \overline{u}_r}{\partial x} + \frac{2\overline{u}_\theta}{r}\hat{u}_\theta  -\frac{\partial \hat{p}}{\partial r}+ \nu_\text{eff} \left[\Delta \hat{u}_r-\frac{2\textrm{i}m}{r^2} \hat{u}_\theta\right] + \hat{f}_r, \label{eqn:LNS2}\\
 -\textrm{i}\omega \hat{u}_\theta & = - \Omega \hat{u}_\theta - \hat{u}_r \frac{\partial \overline{u}_\theta}{\partial r} - \frac{\overline{u}_r}{r} \hat{u}_\theta -\frac{\overline{u}_\theta}{r} \hat{u}_r - \frac{ \textrm{i}m}{r} \hat{p} + \nu_\text{eff} 
 \left[\Delta\hat{u}_\theta+\frac{2\textrm{i}m}{r^2} \hat{u}_r\right] + \hat{f}_\theta, \label{eqn:LNS3}\\
  0 & = \frac{1}{r} \frac{\partial (r \hat{u}_r)}{\partial r}+\frac{\textrm{i}m}{r}  \hat{u}_\theta+\frac{\partial \hat{u}_x}{\partial x}, \label{eqn:LNS4}
\end{align}
with the following shorthand operators:
\begin{align}
\Omega & =\overline{u}_r \frac{\partial}{\partial r}+\frac{\overline{u}_\theta}{r}   \textrm{i} m +\overline{u}_x \frac{\partial}{\partial x}, \\
\Delta & =\frac{1}{r} \frac{\partial}{\partial r}\left(r \frac{\partial}{\partial r}\right)-\frac{m^2+1}{r^2}+\frac{\partial^2}{\partial x^2}.
\end{align}

We utilize the finite difference method to discretize the governing equations. A second-order central differencing scheme is applied to compute all spatial derivatives of the variables in equations \eqref{eqn:LNS1} - \eqref{eqn:LNS4}. Numerical tests demonstrate that this scheme is much more efficient than the previously employed Chebyshev collocation method \citep{gupta2019low,li2024resolvent}, while maintaining the accuracy for wind turbine wake problems.

\begin{figure}
    \centering
    \includegraphics[width=\linewidth]{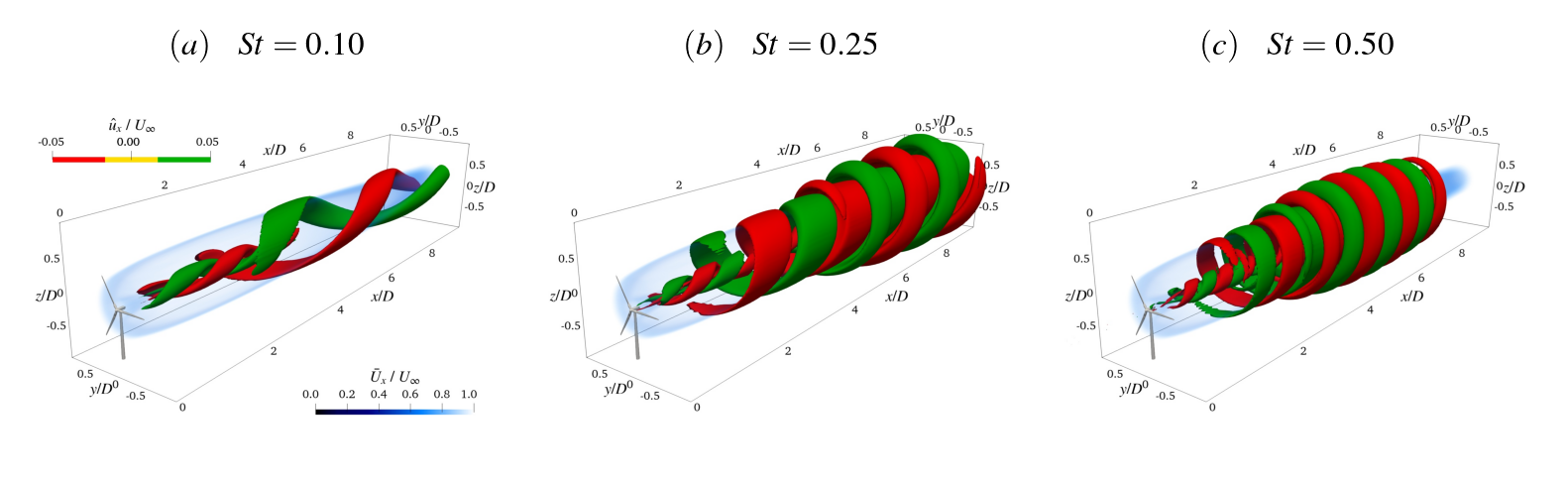}
   
    \caption{ DIPC-induced coherent structures predicted by the linear resolvent model. Helix direction: co-rotating.}
    \label{fig:modelPredictCoherentStructure3D}

\end{figure}

\begin{figure}
    \centering
    \includegraphics[width=\linewidth]{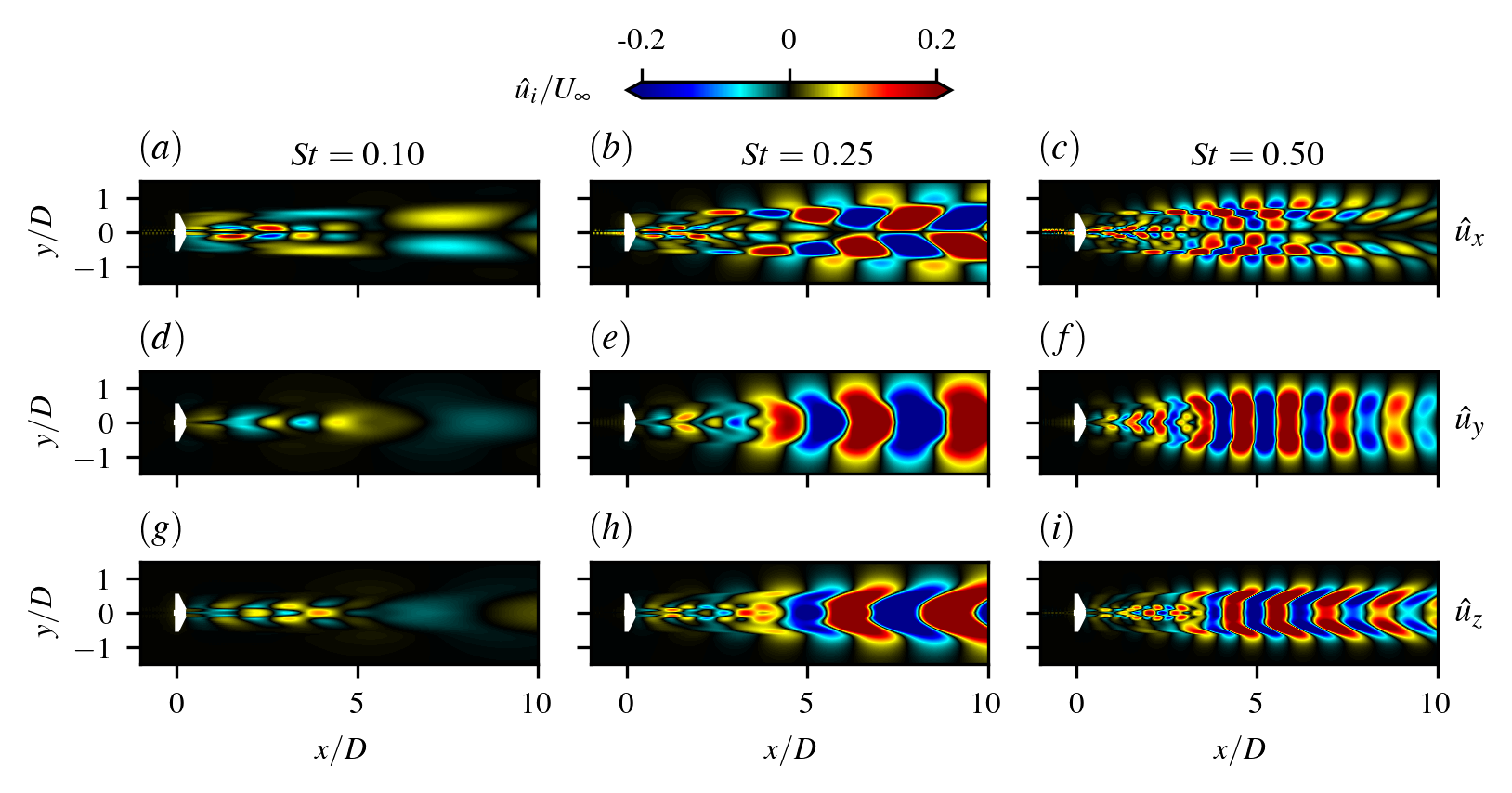}
    \caption{Same as figure \ref{fig:modelPredictCoherentStructure3D}, shown at the hub-height for all three velocity components: streamwise (a-c), lateral (d-f), and vertical (g-i) components, respectively.}
    \label{fig:resolventResponse_helix}
\end{figure}

{\color{black}We employ the resolvent model to predict the linear response of the wake induced by DIPC.} To this end, the resolvent operator is constructed from the time-averaged wake field of the baseline case ($\overline{\boldsymbol{u^0}}$) and the rotor forcing model is employed to compute the perturbation source $\hat{\boldsymbol{f}}$. Note that this is a linear prediction aiming to demonstrate qualitatively correct behaviour rather than precise quantitative results. Results including meanflow modification effects will be presented and validated against LES in \S~\ref{sec:lesValidation}. 

Figure \ref{fig:modelPredictCoherentStructure3D} illustrates three-dimensional coherent structures at different frequencies ($St = 0.10$, $St=0.25$, $St=0.50$) with co-rotating helix. The visualization shows isosurfaces of streamwise velocity fluctuation $\hat{u}_x = \pm 0.05 U_\infty$ in red and green, while the streamwise meanflow velocity is depicted by light blue shadow. The resolvent model, combined with the rotor-forcing model, successfully captures the helical structures and frequency-dependent wake responses. At low frequency ($St = 0.10$), a relatively weak double-helix structure emerges in the wake shear layer, extending from $x = 2D$ to the far wake. The wake response intensifies significantly at $St = 0.25$, characterized by earlier onset of fluctuation in the near wake and larger helical structures in the far wake, albeit with reduced wavelength. At high frequency ($St = 0.50$), the coherent structures exhibit further reduced wavelength, and the double-helix structure ($\hat{u}_x = \pm 0.05 U_\infty$) terminates before $x = 8D$, indicating decreased far-wake fluctuation energy. Additionally, a secondary structure appears behind the hub at all frequencies, but remaining confined to the near wake region.

These wake responses are further investigated in the hub-height plane for all three velocity components in figure \ref{fig:resolventResponse_helix}. On this plane, the three-dimensional coherent structures are reflected by anti-symmetric distribution of $\hat{u}_x$ and symmetric distribution of $\hat{u}_y$ and $\hat{u}_z$, indicating that the helical meandering wake motion is correctly captured by the resolvent model. Again, the wake fluctuation magnitude is the weakest for $St = 0.10$ and becomes stronger for $St=0.25$ and $St=0.50$. The reduction of streamwise wavelength is also observed. While the integrated rotor-forcing and linear resolvent models demonstrate qualitatively accurate wake response predictions, subsequent LES comparisons in \S~\ref{sec:lesValidation} reveal amplitude overestimation. This discrepancy, also observed in motion-to-wake modelling studies \citep{li2024resolvent}, arises from unaccounted mean wake modifications. To address this limitation, later sections will introduce iterative meanflow correction to improve the predictive accuracy of both meanflow and coherent structures.

\subsection{Reynolds stress model \label{sec:RSSModel}}

Reynolds stress (RS) plays a pivotal role in the momentum exchange between the low-speed wake and the high-speed freestream \citep{Gambuzza2023influence, van2024maximizing}, serving as the crucial link between wake fluctuation and mean recovery. Traditional RS modelling based on meanflow gradients proves insufficient for free shear flows containing large-scale coherent structures \citep{Wu2019nonlinear}, because these structures, significantly contributing to RS, are strongly influenced by upstream disturbances and cannot be effectively described by local meanflow strain rate and effective viscosity alone. This additional contribution from coherent fluctuation manifests mathematically as the additional source term ($\nabla \cdot \boldsymbol{\tau}^c$) in the Reynolds-Averaged Navier-Stokes equations for mean velocity field (equation \ref{eqn:meanMom}).

\begin{figure}
    \centering
    \includegraphics[width=\linewidth]{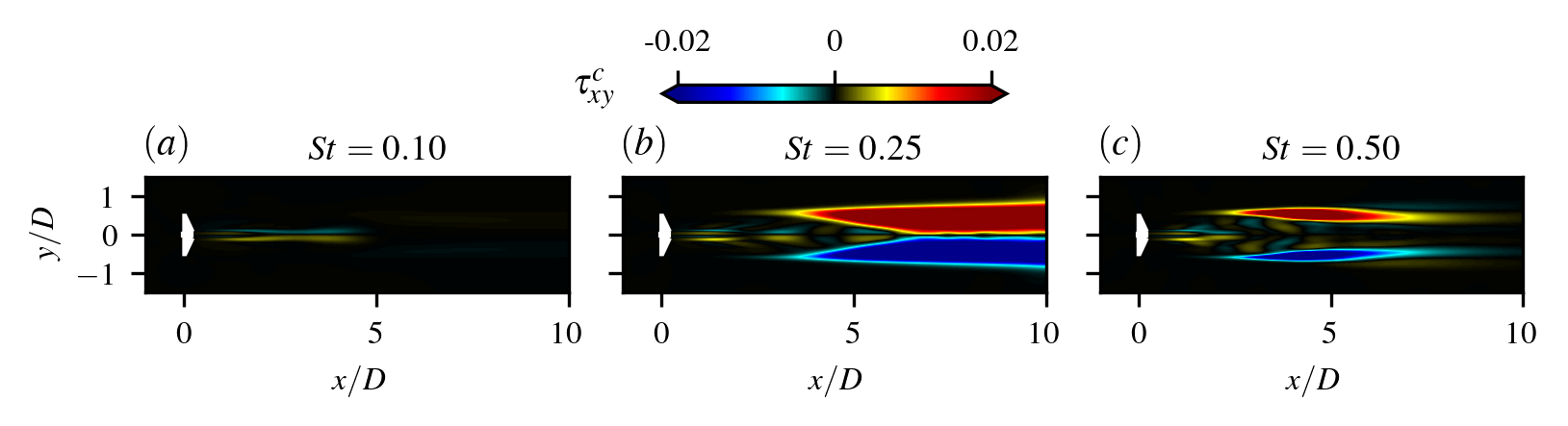}
    \caption{Coherent Reynolds shear stress  $\tau_{xy}^c$ on the hub-height plane for DIPC at different frequencies computed using the linear resolvent model's prediction in figure \ref{fig:resolventResponse_helix}.}
    \label{fig:resolventResponse_RS_helix}
\end{figure}

In this section, we develop a model to predict the RS generated by both coherent wake fluctuation and small-scale turbulence, where the total RS is decomposed into two components:
\begin{equation}
    \boldsymbol{\tau} = \boldsymbol{\tau^{c}} + \boldsymbol{\tau^{s}}. \label{eqn:RSs}
\end{equation}
Here, $\boldsymbol{\tau^{c}}$ represents the contribution from coherent wake fluctuations and $\boldsymbol{\tau^{s}}$ accounts for incoherent small-scale turbulence. The coherent component $\boldsymbol{\tau^{c}}$ is directly computed from the resolvent-predicted wake response $\hat{\boldsymbol{u}}$ through the time averaging of velocity fluctuation products:
\begin{equation}
\begin{split}
    \tau^{c}_{ij} & = - \overline{u^c_i u^c_j} \\ 
    & = -\overline{ \frac{1}{2}\left(\hat{u}_i e^{-\text{i}\omega t} +  \hat{u}_i^H e^{\text{i}\omega t} \right) \cdot \frac{1}{2}\left(\hat{u}_j e^{-\text{i}\omega t} +  \hat{u}_j^H  e^{\text{i}\omega t} \right)}   \\ 
    & = -\frac{1}{4} \left(\hat{u}^{H}_i \hat{u}_j + \hat{u}^{H}_j \hat{u}_i \right) \\
    & = -\frac{1}{2} \mathcal{R} \left( \hat{u}^{H}_i \hat{u}_j \right),
 \end{split}\label{eqn:RSC2}
\end{equation}
where $\mathcal{R}$ denotes the real part of a complex number {\color{black}and $\{\cdot\}^H$ denotes the complex conjugate operator}. Figure \ref{fig:resolventResponse_RS_helix} shows the coherent component of the $\tau_{xy}^c$ on the hub-height plane using coherent structures shown in figure \ref{fig:resolventResponse_helix}. As seen, the $\tau_{xy}^c$ for different Strouhal numbers shows distinct spatial patterns and strength, indicating that the model predicts different momentum flux contributed by the coherent structures at different Strouhal numbers.  At $St = 0.10$, the RS remains relatively weak and spatially confined to the centerline of the near wake as a result of the weak coherent structures shown in figures \ref{fig:resolventResponse_helix} (a) and (d). As the Strouhal number increases to $St = 0.25$, the coherent structures generate pronounced $\tau_{xy}^c$ extending to the far wake. The region of elevated $\tau_{xy}^c$ shifts toward the wake edge rather than remaining at the wake centre, in contrast to the $St=0.10$ case. At $St = 0.50$, both the magnitude and spatial extent of the $\tau_{xy}^c$ diminish compared to $St = 0.25$, characterized by a notable decrease beyond the intermediate wake region $(x>5D)$. This reduction implies an attenuation of coherent velocity fluctuations in the far field at $St=0.50$, which is also observed in figures \ref{fig:resolventResponse_helix} (c) and (f). Again, it is worth noting that the result presented in figure \ref{fig:resolventResponse_RS_helix} is only a qualitative prediction and may deviate from reality, since no mean wake correction is taken into account. 

The second component of $\tau$, i.e., the contribution from incoherent small-scale turbulence, is modelled using the Boussinesq assumption, which introduces an effective eddy viscosity $\nu_\text{eff}$:
\begin{equation}
    \tau^{s}_{ij} =   \nu_\textrm{eff} \frac{\partial \overline{u}_i}{\partial x_j}. \label{eqn:RSs2}
\end{equation}
It is then connected to the meanflow equation \eqref{eqn:meanMom} by assuming  $\nu_\text{eff} \nabla^2 \overline{\boldsymbol{u}} \approx \boldsymbol{\nabla \cdot \tau^s}$. To determine $\nu_\text{eff}$, we utilize data from the uncontrolled baseline wake, where the RS arises solely from small-scale turbulence, with the following equation:
\begin{equation}
- \overline{u_x' u_y'} = \nu_\text{eff} \frac{\partial \overline{u_x}}{\partial y}. \label{eqn:fitNut}
\end{equation}
Here, $-\overline{u_x' u_y'}$ denotes the hub-height in-plane primary shear stress and $\displaystyle \frac{\partial \overline{u_x}}{\partial y}$ represents the meanflow strain rate, both obtained from the baseline wake. The effective viscosity $\nu_\text{eff}$ is then computed using a least-squares method at each downstream location. For prediction purposes, this effective viscosity, once determined, remains fixed and is applied to all controlled wake cases.  The validity of this assumption will be assessed using large-eddy simulation results in $\S$ \ref{sec:lesResult}.

\subsection{Meanflow model \label{sec:meanFlowCorrModel}}

The meanflow model proposed here predicts the enhanced wake recovery induced by AWC, serving as a critical component in the nonlinear iterative framework (figure \ref{fig:modellingFramework}). Although equations \eqref{eqn:meanMom} and \eqref{eqn:meanIncom} allow direct solution of the time-averaged wake from RS, we adopt a more efficient approach by leveraging the known uncontrolled baseline wake. Specifically, using the RS predictions from \S \ref{sec:RSSModel} as input, the model computes a correction to the baseline flow. This correction-based formulation expresses the modified time-averaged wake as:

\begin{align}
    \overline{\boldsymbol{u}} & = {\overline{\boldsymbol{u}^0}} + {\delta \boldsymbol{u}}  \label{eqn:meanVelocityUpdate},   \\
        \overline{p} & = \overline{p^0} + {\delta p},  \\
            \overline{\boldsymbol{f}} & = {\overline{\boldsymbol{f}^0}} + {\delta \boldsymbol{f}}  \approx \overline{\boldsymbol{f}^0}, 
\end{align}
where $[\overline{\boldsymbol{u}}, \overline{p}, \overline{\boldsymbol{f}}]$ represent the time-averaged velocity, pressure, and rotor forcing of the corrected case, $\{\cdot\}^0$ denotes the result of baseline wake, and $\delta \{\cdot\}$ denotes the modifications. We assume the time-averaged rotor forcing modification is negligible  ($\delta \boldsymbol{f} \approx 0$), which is consistent with the rotor-forcing model developed based on linear assumption. 

Assuming that the uncontrolled wake satisfies the following governing equations:

\begin{align} 
 \left({\overline{\boldsymbol{u^{0}}}} \cdot \nabla \right) \overline{\boldsymbol{u^{0}}}
 + \nabla \overline{p^0} - \nu_\text{eff} \nabla^2 \overline{\boldsymbol{u^{0}}} & = \overline{\boldsymbol{f^{0}}}, \label{eqn:meanBaseMom} \\
 \nabla \cdot \overline{\boldsymbol{u^{0}}} & = 0 \label{eqn:meanBaseIncom}.
\end{align}
Subtracting them from the meanflow governing equations \eqref{eqn:meanMom} and \eqref{eqn:meanIncom} results in the governing equations for the correction variables: 
\begin{align} 
 \left(\overline{\boldsymbol{u^0}} \cdot \nabla \right) \delta \boldsymbol{u} +  \left({\delta \boldsymbol{u}} \cdot \nabla \right) \overline{\boldsymbol{u^0}} + 
 \left({\delta \boldsymbol{u}} \cdot \nabla \right) \delta \boldsymbol{u} 
 + \nabla ({\delta p}) - \nu_\text{eff} \nabla^2 \delta \boldsymbol{u}& =  \nabla \cdot \boldsymbol{\tau^c} , \label{eqn:meanCorrMom} \\
 \nabla \cdot \delta \boldsymbol{u} & = 0 \label{eqn:meanCorrIncom},
\end{align}
with $\delta \boldsymbol{u}$ and $\delta  p$ the unknowns to be solved and $\nabla \cdot \boldsymbol{\tau^c}$ the source term driving the wake modification. These equations \eqref{eqn:meanCorrMom}-\eqref{eqn:meanCorrIncom} are also solved in a two-dimensional $(x,r)$ domain with azimuthal uniformity, given the statistically axisymmetric nature of the forcing and responses.  The equations are discretized using a second-order central differencing scheme. The pressure-velocity coupling is solved via the projection method \citep{chorin1967numerical}. The nonlinear  term $ \left({\delta \boldsymbol{u}} \cdot \nabla \right) \delta \boldsymbol{u}$ is treated through inner iterations.

\subsection{Iterative solution for self-consistent prediction \label{sec:iterativescheme}}

\begin{table}
\centering  
\caption{The iterative procedure for computing self-consistent wake response under AWC. \label{tab:algorithm} }
\begin{tabular}{p{0.8\textwidth}}  
\hline
\begin{itemize}  
    \item Set initial condition $\overline{\boldsymbol{u}} = {\overline{\boldsymbol{u^0}}} $ and $n = 0$ 
    \item Start nonlinear iteration with a small forcing amplitude $A_0 = A_f/N$   
    \begin{enumerate}
        \item Compute the rotor forcing ($\S$ \ref{sec:forcingModel}): $A_n \rightarrow \hat{\boldsymbol{f}} $  
        \item Solve wake fluctuation by the resolvent model  ($\S$ \ref{sec:resolventModel}):  $\hat{\boldsymbol{f}} \rightarrow \hat{\boldsymbol{u}}$  
        \item Calculate Reynolds stress forcing ($\S$ \ref{sec:RSSModel}): $\frac{1}{2}\mathcal{R}((\hat{\boldsymbol{u}}^H \hat{\boldsymbol{u}}) \rightarrow \boldsymbol{\tau^c} $  
        \item Solve meanflow correction ($\S$ \ref{sec:meanFlowCorrModel}), $\boldsymbol{\tau^c} \rightarrow \delta{u}_n$  
        \item Calculate the new meanflow $\overline{\boldsymbol{u}}  =\overline{\boldsymbol{u^0}} + \delta \boldsymbol{u}_n $ 
        \item $n+1 \rightarrow n$
        
        \begin{itemize} 
            \item   If $A_n < A_{f}$, increase $A_n \leftarrow A_{n-1} + A_f/N  $ and go to step (i) 
            \item  If $A_n = A_{f}$, go to step (i) and repeat until {\color{black}the relative step difference of meanflow modification is smaller than the  tolerance} 
            \begin{equation*}
              \frac{|\delta \boldsymbol{u}_n - \delta \boldsymbol{u}_{n-1} |}{|\delta \boldsymbol{u}_n|}   < \epsilon
            \end{equation*}
        \end{itemize}

    \end{enumerate}  
        \item Save converged self-consistent prediction of $\hat{\boldsymbol{u}}$  and $\overline{\boldsymbol{u}}$.
\end{itemize} 
\end{tabular}  
\end{table} 

The submodels introduced above are integrated and solved through an iterative procedure, as illustrated in figure \ref{fig:modellingFramework} and detailed in table \ref{tab:algorithm}. Beginning with a user-defined AWC and an uncorrected baseline wake, the model sequentially computes the rotor forcing ($\hat{\boldsymbol{f}}$), the coherent structures ($\hat{\boldsymbol{u}}$), the Reynolds stress ($\boldsymbol{\tau}$), and the meanflow correction ($\delta \boldsymbol{u}$). After obtaining $\delta \boldsymbol{u}$, the time-averaged wake $\overline{\boldsymbol{u}}$ is updated through equation \eqref{eqn:meanVelocityUpdate}, serving as the baseflow for the subsequent iteration's resolvent analysis. 

The solution naturally converges through an inherent self-correcting mechanism. For instance, an overestimated velocity fluctuation $\hat{\boldsymbol{u}}$ generates a larger $\delta \boldsymbol{u}$, which weakens the wake deficit in the subsequent iteration, thereby reducing $\hat{\boldsymbol{u}}$; the converse also holds true. To ensure numerical stability, the AWC amplitude is gradually increased from an initial value $A_0$ to the target amplitude $A_f$ over $N = 5$ iterations, after which it remains constant at $A_f$. Convergence is monitored through the relative variation of meanflow modification $\delta \boldsymbol{u}$ between consecutive steps:
\begin{equation}
    \frac{|\delta \boldsymbol{u}_n - \delta \boldsymbol{u}_{n-1} |}{|\delta \boldsymbol{u}_n|}  < \epsilon, \label{eqn:residual}
\end{equation}
where the tolerance $\epsilon = 10^{-2}$ is selected to balance between computational efficiency and solution accuracy, {\color{black}which is discussed in appendix \ref{sec:stoppingcriteria}. }

 \begin{figure}
    \centering
    \includegraphics[width=\linewidth]{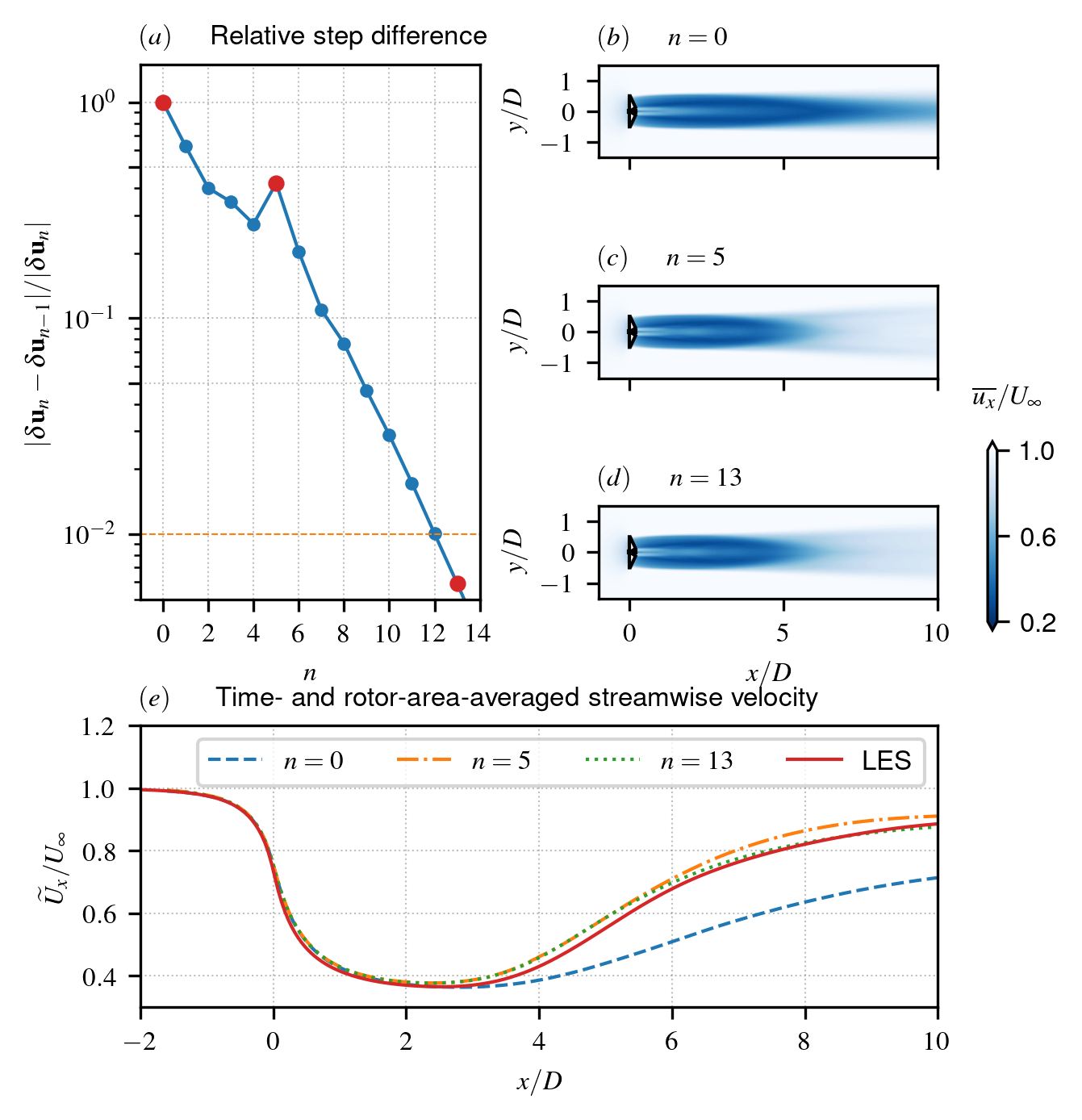}
    \caption{Convergence characteristics and wake modification predicted by the proposed model: (a) convergence history showing {\color{black}the relative step difference of the meanflow modification}; (b-d) hub-height contours of time-averaged streamwise velocity at iterations $n=0$ (initial state), $n=5$ (intermediate), and $n=13$ (converged solution), respectively;  (e) streamwise evolution of time- and rotor-area-averaged streamwise velocity, comparing present predictions with large-eddy simulation results.  The AWC approach is DIPC with $St=0.25$ and co-rotate helix.}
    \label{fig:IterativeResults}
\end{figure}

The nonlinear solution procedure and convergence characteristics are demonstrated in figure~\ref{fig:IterativeResults} for the co-rotating DIPC strategy and forcing frequency $St=0.25$. The convergence history, monitored through the {\color{black}relative step difference} defined by equation \eqref{eqn:residual}, exhibits nearly monotonic decay from unity to below $10^{-2}$ over 14 iterations (figure \ref{fig:IterativeResults} a). This robust convergence is interrupted only once at $n=5$, corresponding to the transition from the initial to the target forcing amplitude $A_f$. Beyond this point, the {\color{black}relative step difference} decreases exponentially until reaching the threshold $\epsilon = 10^{-2}$ at iteration $n = 13$.

The evolution of the hub-height streamwise velocity field $\overline{u}_x$ is illustrated in figures~\ref{fig:IterativeResults} (b-d). Starting from the unmodified initial state ($n=0$), the wake undergoes substantial expansion by iteration $n=5$, with subsequent iterations refining the solution until convergence at $n=13$. This progression demonstrates the model's capability to capture the AWC-induced wake recovery through the nonlinear iterative procedure.

Figure \ref{fig:IterativeResults} (e) plots the streamwise evolution of time- and rotor-area-averaged streamwise velocity ($\widetilde{U}_x$) and provides quantitative validation against LES results obtained for the same scenario. In the near-wake ($0 <x < 2D$), all curves collapse, indicating a decrease of $\widetilde{U}_x$ to $0.4U_\infty$ that matches well with LES predictions. The intermediate wake region ($2D < x < 6D$) shows the onset of improved wake recovery by LES, which is successfully predicted by the nonlinear solutions at $n=5$ and $n=13$. The amount of velocity increase reaches approximately $0.2U_\infty$ at $x=6D$. In the far-wake ($x > 6 D$), this velocity difference between the controlled and uncontrolled cases ceases to grow. At $x = 10D$, the controlled wake reaches approximately 90\% of freestream wind speed. The converged solution ($n=13$) shows the optimal agreement with LES, while the intermediate solution ($n=5$) slightly overpredicts wake velocity. Overall, the good agreement between the present method and LES results suggests that the proposed model successfully captures the essential physics of wake development.

\section{Model validation against LES \label{sec:ModelValidation}}

The model's predictive capability is validated by LES in this section. First, \S~\ref{sec:LES} introduces the LES method, the wind turbine model, the parameter space of AWC, and the numerical setup employed. Next, $\S$ \ref{sec:lesResult} provides an investigation of the AWC-induced wake response by analysing the LES results, with special attention paid to the coupling between the coherent structures and time-averaged wake. In $\S$ \ref{sec:lesValidation}, a detailed comparison between the model and LES results will be provided to demonstrate the predictive capability of the proposed model.

\subsection{LES method and configuration \label{sec:LES}}

We simulate the fully nonlinear evolution of wind turbine wakes with the large eddy simulation code VFS-Wind \citep{yang2015VFS}, which has been validated and applied to study the fluid mechanisms for wind turbine wake evolution \citep{yang2016coherent,foti2018similarity,li2021large,dong2023bladeDesign, li2024resolvent}.

The code implements the filtered incompressible Navier-Stokes equations to solve the three-dimensional turbulent airflow. The Coriolis force is not considered to simplify the problem. The dynamic Smagorinsky model \citep{Smagorinsky,germano1991subGridModel} is employed for subgrid-scale stress modelling. A well-validated actuator surface method (ASM, \cite{yang2018ASMethod}) is employed to model the rotor and nacelle, which are both represented by equivalent aerodynamic forces. The smoothed discrete delta function \citep{yang2009Kernel} is employed to transfer the aerodynamic forces to the computational grid to avoid singularity issues. The present work employs a structured Cartesian grid for discretizing the computational domain,  a second-order central differencing scheme for spatial discretization, and a second-order fractional step method for temporal integration. A detailed description of the flow solver can be found in \citep{ge2007numerical}.

\subsubsection{Wind turbine model}

The simulations employ the International Energy Agency's 10 MW horizontal-axis reference wind turbine \citep{iea2019}. This turbine has a three-blade rotor with a diameter of $D=198$ m. The nacelle is mimicked by a cylinder measuring 10 m in width and 5 m in length. For the sake of simplicity, the tower has been omitted and the blades are assumed to be rigid in the simulations.   

In the simulation, the wind turbine operates near its design state, rotating in the clockwise direction at a fixed tip speed ratio of $\lambda = 9$ with a uniform incoming wind speed of $U_\infty = 10~\text{m/s}$.  In this scenario, the LES predicts a time-averaged thrust coefficient of $C_T = 0.75$ and a power coefficient of $C_P = 0.48$, being close to the rotor’s design state.

\subsubsection{Parameter space of LES}


Table \ref{tab:uniformInflowParamSpace} presents the parameter space for the LES. In total 25 large eddy simulations are carried out, including one reference baseline case without AWC and 24 cases with three AWC techniques, four Strouhal numbers, and two helix directions. The control amplitude is kept small to avoid disturbing wind turbine operation \citep{wei_dabiri_2023}, yet sufficient to generate noticeable wake response and recovery within sensitive frequency ranges \citep{cheung2024fluid}. The translation and precession amplitudes are equivalent in terms of blade tip displacement: $A_\gamma R = 0.01D = A_t $. The selected Strouhal numbers span the wake's characteristic frequency range. Among them, $St=0.25$ and $St=0.35$ are close to the optimal forcing frequency for DIPC, according to the large eddy simulation of \cite{frederik2020helix}. The other two frequencies, $St = 0.10$ and $St=0.50$, are below and above the wake's most sensitive frequency, as determined by linear resolvent analysis in a previous study on the same wind turbine's wake \citep{li2024resolvent}.

\begin{table}
\begin{center}
\begin{tabular}{cccc}
\def~{\hphantom{0}}
      & ~~~~~~~~~~~~~~Amplitude~~~~~~~~~~~~~~    & Strouhal Number       & Helix direction     \\[3pt]
Ref & -            & - & -                          \\[3pt]
DPIC & $A_\beta = 0.5 ^\circ $ & $St \in\{0.10, 0.25, 0.35, 0.50\}$  & Co \& Counter rotation\\[3pt]
Translation  & $A_t =0.01D$   & $St \in\{0.10, 0.25, 0.35,  0.50\}$  & Co \& Counter rotation\\[3pt]
Precession  & $ A_\gamma = 0.02$ rad & $St \in\{0.10, 0.25, 0.35, 0.50\}$  & Co \& Counter rotation                
\end{tabular}

\caption{Parameter space of LES investigation.}
\label{tab:uniformInflowParamSpace}
\end{center}
\end{table}

\subsubsection{Numerical configurations}

\begin{figure}
    \centering
    \includegraphics[width=\textwidth]{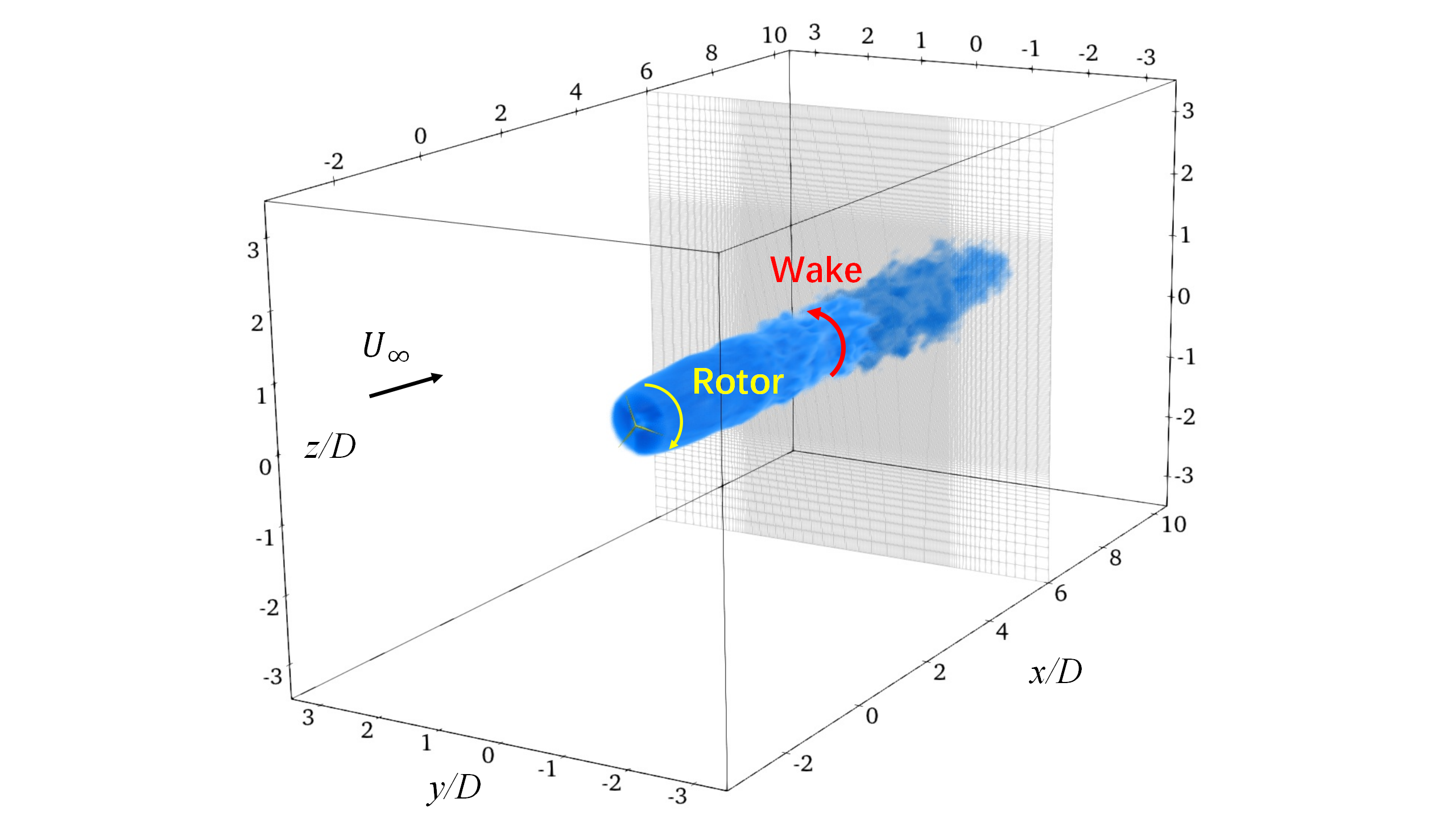}
    \caption{Schematic of the computational domain and the grid for LES. }
    \label{fig:computationalDomain}
\end{figure}

The computational domain is rectangular, as depicted in figure~\ref{fig:computationalDomain}, extending $14D$ in the streamwise direction ($x$) and $7D$ in both transverse ($y$) and vertical ($z$) directions.  The wind turbine is positioned on the domain centerline,  $3.5D$ downstream of the inlet boundary, with the hub centre defining the origin of the coordinate system.  The {\color{black} slip boundary} condition is applied to the four lateral boundaries. At the inlet, a uniform inflow of $U_\infty = 10$ m/s is imposed, neglecting the effects of wind shear, veer, turbulence, and the ground. At the outlet, the Neumann boundary condition is applied to the velocity components. The grid interval is $\Delta x = D/20$ and $\Delta y = \Delta z = D/40$ in the wake region and the time step is controlled so that the blade does not pass through more than one grid per step. 

\subsection{LES results\label{sec:lesResult}}

This section analyses the impact of AWC on coherent structures and their role in mean wake recovery by analysing LES data, to obtain physical insight and to verify important assumptions of the wake modelling.     

We begin by examining the instantaneous helical wakes produced by the three different rotor actuation approaches in \S~\ref{sec:InstantanousWake}, followed by an exploration of the coherent structures in \S~\ref{sec:coherentstructures} and their effect on enhanced wake recovery in \S~\ref{sec:Time-averagedWake}. These analyses highlight the similarities and differences caused by various AWC configurations.  Special emphasis is placed on the RS in \S~\ref{sec:RSS}, which is an essential element to relate AWC-induced coherent structures and recovery for achieving self-consistent wake prediction in the proposed model. Only the results at $St = 0.10$, 0.25, and 0.50 are presented in detail, because the results at $St = 0.35$ closely resemble those of $St = 0.25$ and are reserved only for validating the proposed wake model in §~\ref{sec:lesValidation} to avoid redundancy.

\subsubsection{Instantaneous wake \label{sec:InstantanousWake}}

Figure \ref{fig:InstantanousWake} provides a three-dimensional visualisation of instantaneous wind turbine wakes under various AWC strategies at a forcing frequency of $St=0.25$. Despite the differences in AWC methods, the wake structures demonstrate notable similarity across cases, characterised by the downstream amplification of perturbations due to convective instability \citep{gupta2019low}. Near the rotor, the wake closely resembles the baseline case; however, further downstream, the helical structure amplifies significantly, causing the wake to deflect from the centre line and form a pronounced meandering pattern. These characteristics align with those observed in DIPC-controlled wakes in earlier studies \citep{frederik2020helix,korb2023characteristics}, regardless of the specific AWC techniques applied.

\begin{figure}
    \centering
\includegraphics[trim={0cm 6cm 0cm 0cm}, clip, width=\textwidth]{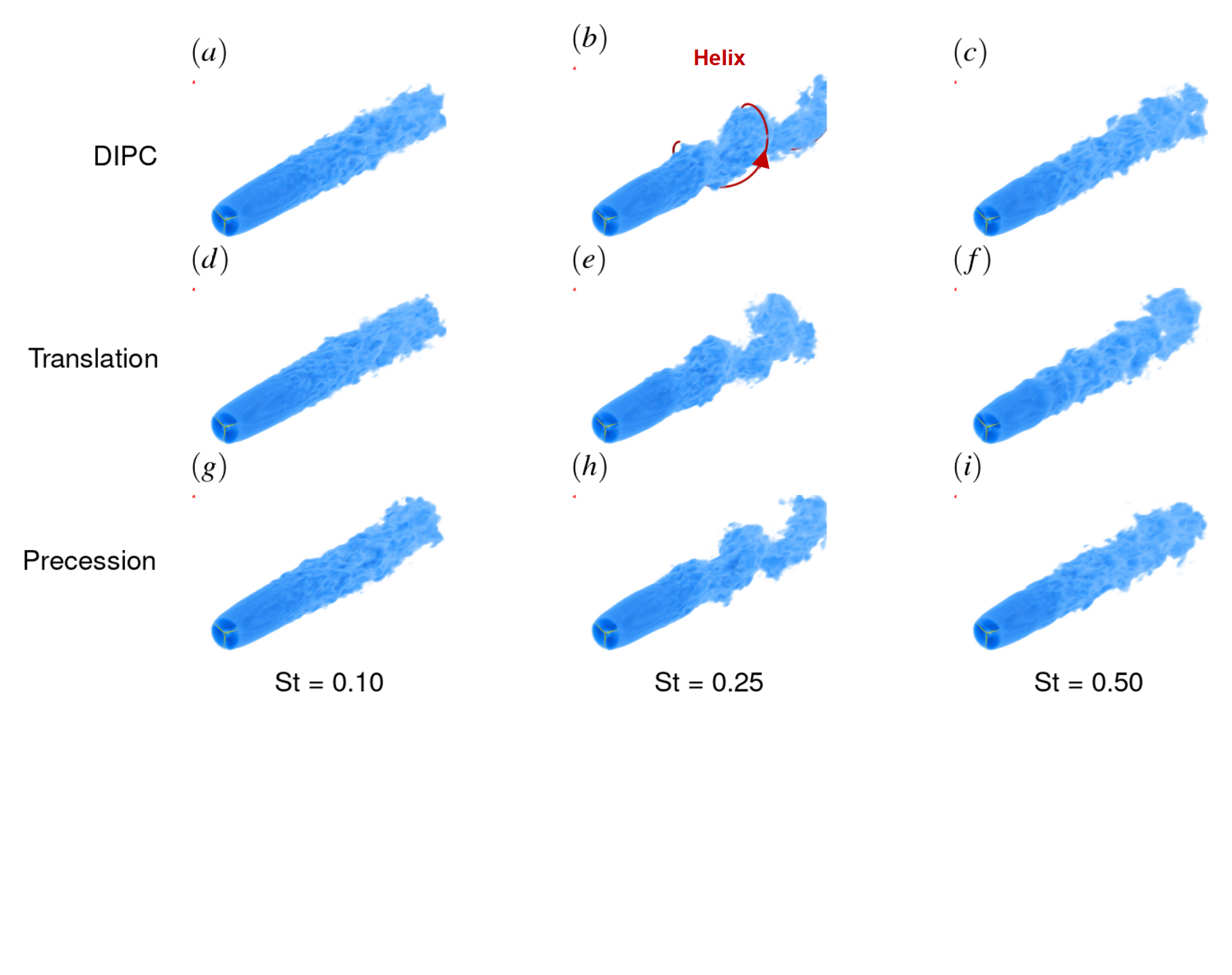}  
    \caption{Instantaneous wind turbine wake forced by DIPC (a)(b)(c), translation (d)(e)(f) and precession (g)(h)(i) at different frequencies, St = 0.10~(a)(d)(g); St = 0.25~(b)(e)(h), St= 0.50~(c)(f)(i). Helix direction: co-rotating.}
    \label{fig:InstantanousWake_frequencyCompare}
\end{figure}

The effect of excitation frequency on the wake structures under different AWC techniques is illustrated in figure \ref{fig:InstantanousWake_frequencyCompare}. The instantaneous wakes exhibit consistent frequency-dependent trends across all AWC approaches. At the lowest frequency ($St=0.10$), the impact of AWC is minimal, with the wake showing little deformation (figures \ref{fig:InstantanousWake_frequencyCompare} a, d, and g). As the frequency increases to $St=0.25$, significant amplification of the wake response occurs, producing pronounced large-scale helical deformations (figures \ref{fig:InstantanousWake_frequencyCompare} b, e, and h). However, further increasing the frequency to $St=0.50$ suppresses these large-scale deformations in the far wake, leading instead to the emergence of small-scale wavy deformations (figures \ref{fig:InstantanousWake_frequencyCompare} c, f, and i).

These observations indicate that while initial disturbances differ among AWC techniques, the frequency-dependent behaviour of downstream wake evolution remains qualitatively similar. The reason is that the sensitivity of the wake to specific excitation frequencies is governed by the wake's intrinsic properties, unaffected by the choice of AWC. The primary effect of different AWCs is to induce different rotor forcing so that the magnitude of the wake response can vary quantitatively.


\subsubsection{Coherent structures \label{sec:coherentstructures}}


\begin{figure}
    \centering
    \includegraphics[width=\linewidth]{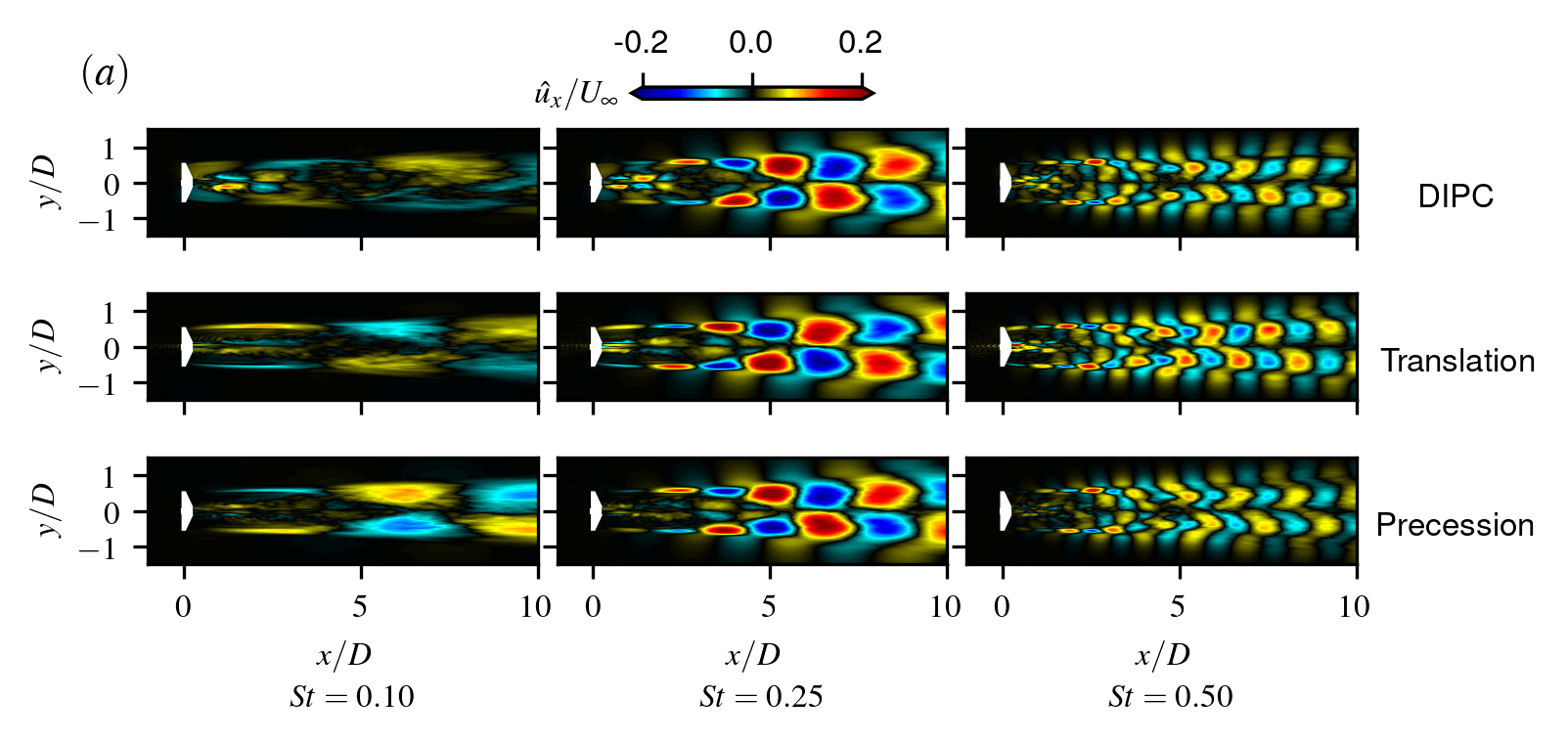}
    \includegraphics[width=\linewidth]{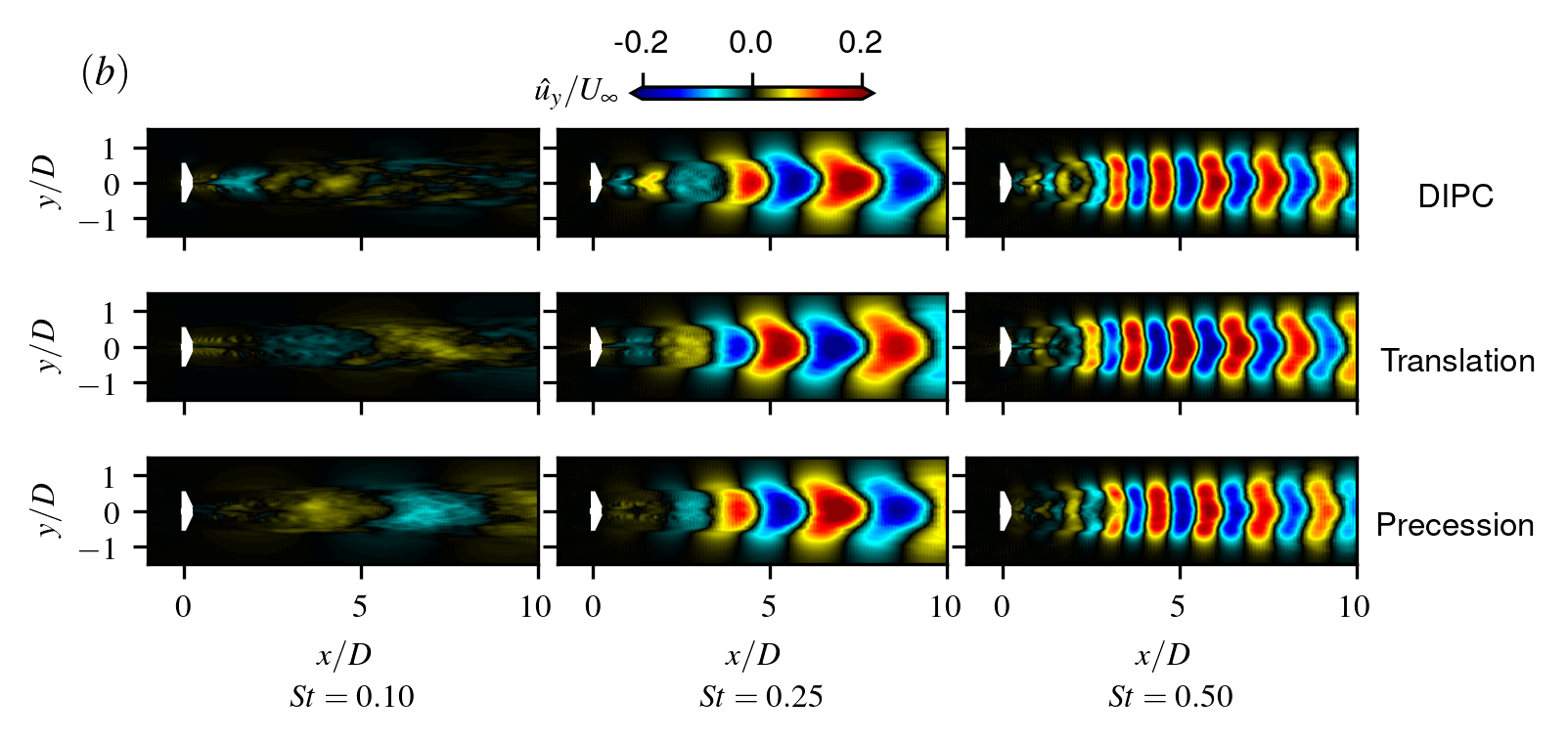}
    \includegraphics[width=\linewidth]{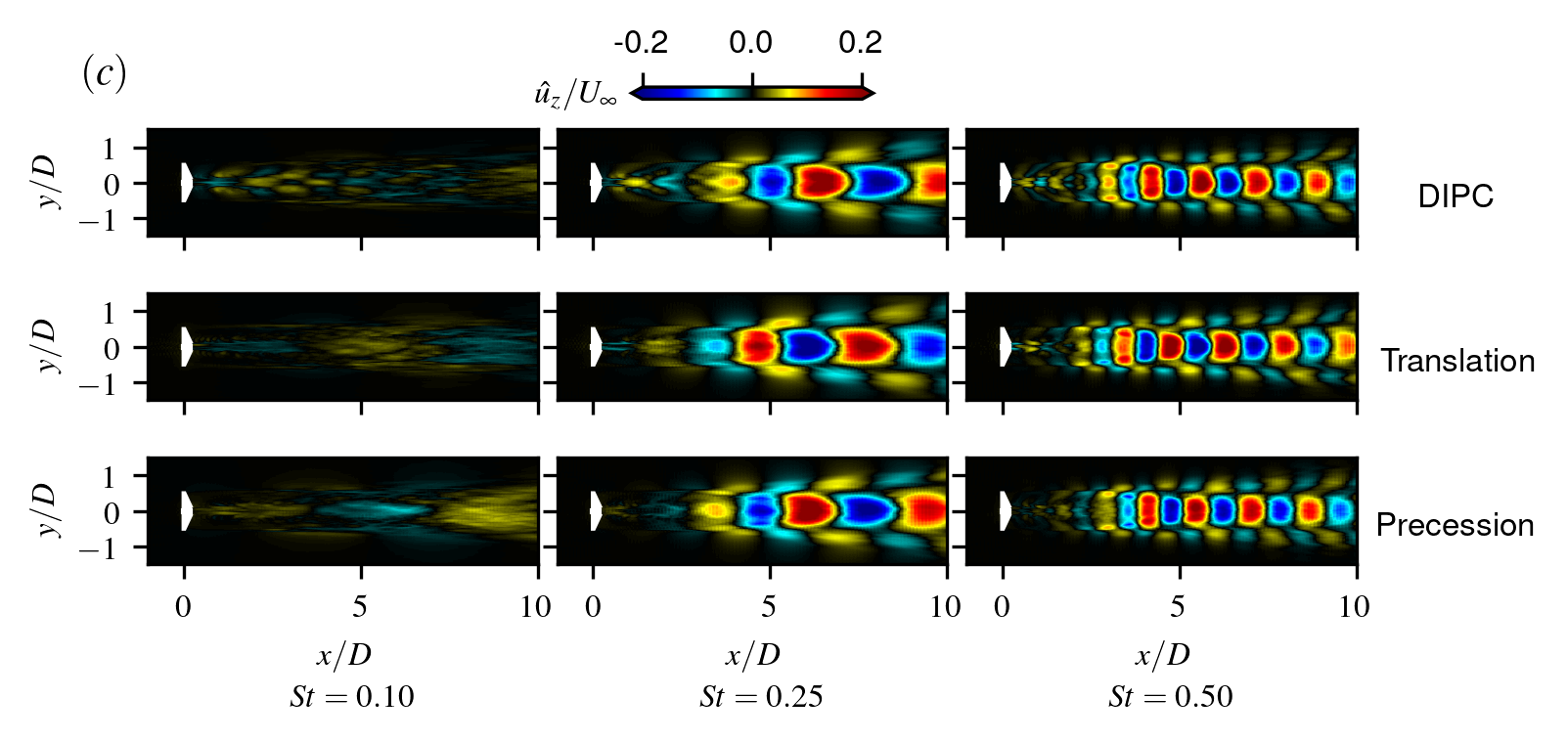}
    \caption{Coherent structures derived from Fourier transform for various AWC approaches at different excitation frequencies. Streamwise velocity (a), horizontal velocity (b), and vertical velocity components (c). Each row represents the same AWC approach, while each column corresponds to the same forcing frequency. Helix direction: co-rotating.}
    \label{fig:FFTVelocity_hubplane}
\end{figure}

To understand in depth the difference made by different AWC, the coherent structures in the wake are analysed. {\color{black} We employ Fast Fourier Transform (FFT) to transform the velocity field from the time domain to the frequency domain. Then,  the spatial mode at the forcing frequency on the hub-height horizontal plane ($xOy$) is extracted, since previous studies have shown that the wake dynamics are locked to the excitation frequency \citep{messmer2024enhanced}.}  This method facilitates the identification of distinct spatial patterns associated with various AWC techniques and excitation frequencies, enabling detailed examination of the similarities and differences among AWC approaches.

In figure \ref{fig:FFTVelocity_hubplane} (a) the structure of the streamwise velocity fluctuation is found to be quasi-anti-symmetric for all considered AWC approaches. This pattern indicates that acceleration on one side of the wake is always accompanied by deceleration on the other side, reflecting the phenomenon that the entire instantaneous wake meanders laterally \citep{korb2023characteristics}. In figures \ref{fig:FFTVelocity_hubplane} (b) and (c), both horizontal and vertical velocities show a symmetric pattern extending laterally in the wake region, indicating that the entire wake is deflected in the directions perpendicular to the streamwise direction. For all velocity components, the fluctuation alternates in sign as travelling downstream, demonstrating the wake’s meandering behaviour. The wavelength of the coherent structures is inversely related to the actuation frequency. The unstable nature of the wake \citep{iungo2013linear,gupta2019low} is revealed by the downstream growth of velocity fluctuation magnitude. The frequency-dependent amplification observed in figure \ref{fig:InstantanousWake_frequencyCompare} is clearly illustrated by the magnitude of velocity fluctuations. The streamwise velocity fluctuation is strongest at $St=0.25$ and weaker at $St=0.50$ and $St=0.10$ across all AWC approaches. For the lateral velocity components, the fluctuations at $St=0.50$ are comparable in magnitude to those at $St=0.25$ but are segmented into shorter structures, explaining the smaller deformation seen in the instantaneous wake in figure \ref{fig:InstantanousWake_frequencyCompare}. 

Qualitative differences across AWC approaches are observed primarily in the vicinity behind the rotor and the near wake. These differences diminish in the far wake, where the coherent structures exhibit qualitative similarity across all AWC approaches. This similarity in the far wake implies a filter effect of the wake which selectively amplifies certain modes determined by its inherent instability properties. Nevertheless, distinct AWC approaches lead to varying fluctuation amplitudes, indicating quantitative differences in their wake modulation effects. The precession approach is found to trigger the strongest fluctuations at $St = 0.10$, whereas the translation approach results in slightly stronger fluctuations at $St=0.50$. At $St = 0.25$, all three AWC approaches yield comparable results in the far wake. Another notable distinction observed in figure \ref{fig:FFTVelocity_hubplane} is the systematic phase variation across AWC techniques, which is caused by the specific implementation of each AWC technique, as detailed in \S~\ref{sec:forcingModel}.

\subsubsection{Wake Recovery \label{sec:Time-averagedWake}}

\begin{figure}
    \centering
    \includegraphics[width=\textwidth]{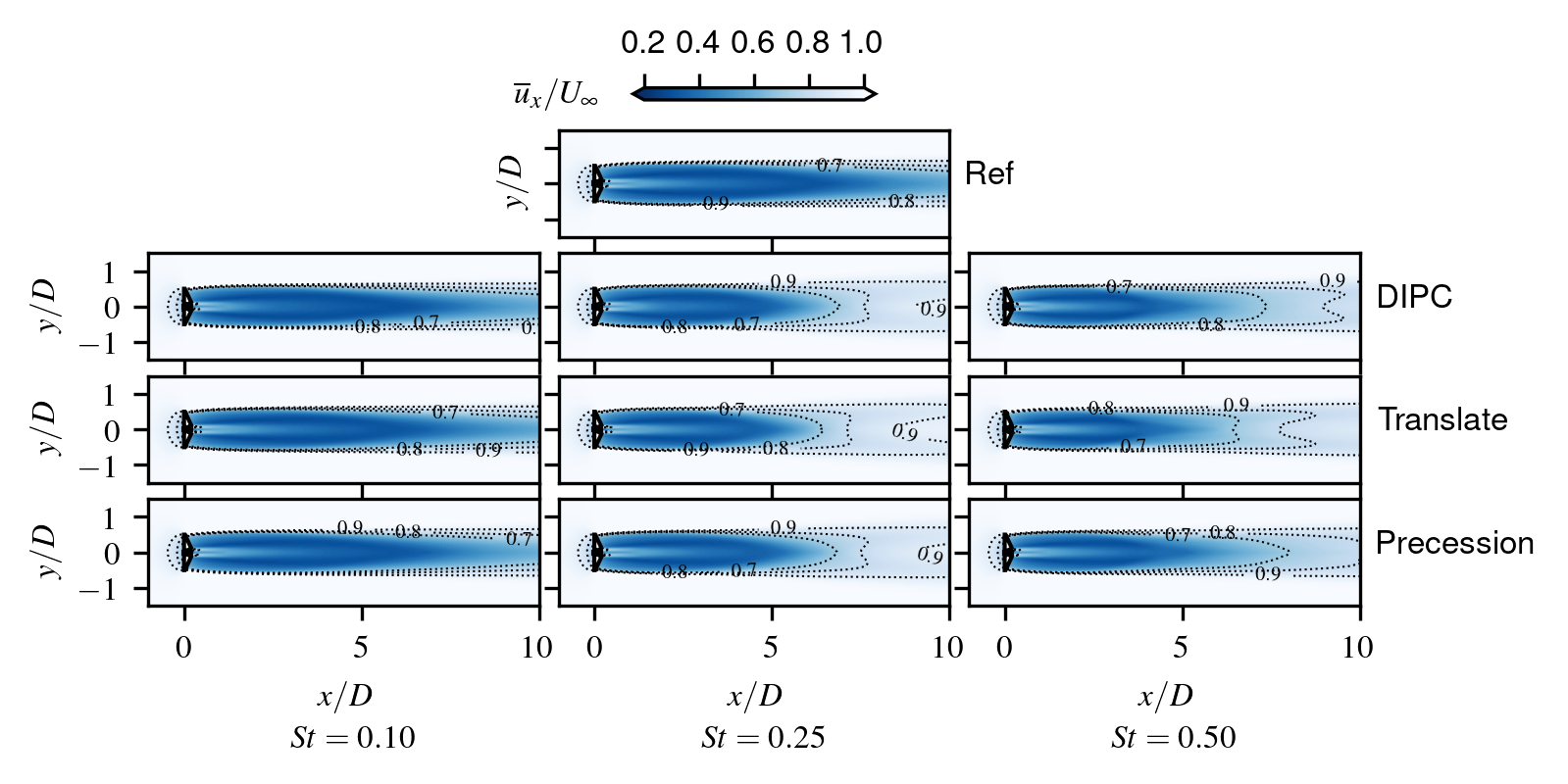}
    \caption{Contour of the time-averaged wind speed for different AWC approaches, forcing frequencies. Helix direction: co-rotation.  The Ref case shows the baseline wake without AWC.}
    \label{fig:timeAveragedWakeContour}
\end{figure}

\begin{figure}
    \centering
    \includegraphics[width=\textwidth]{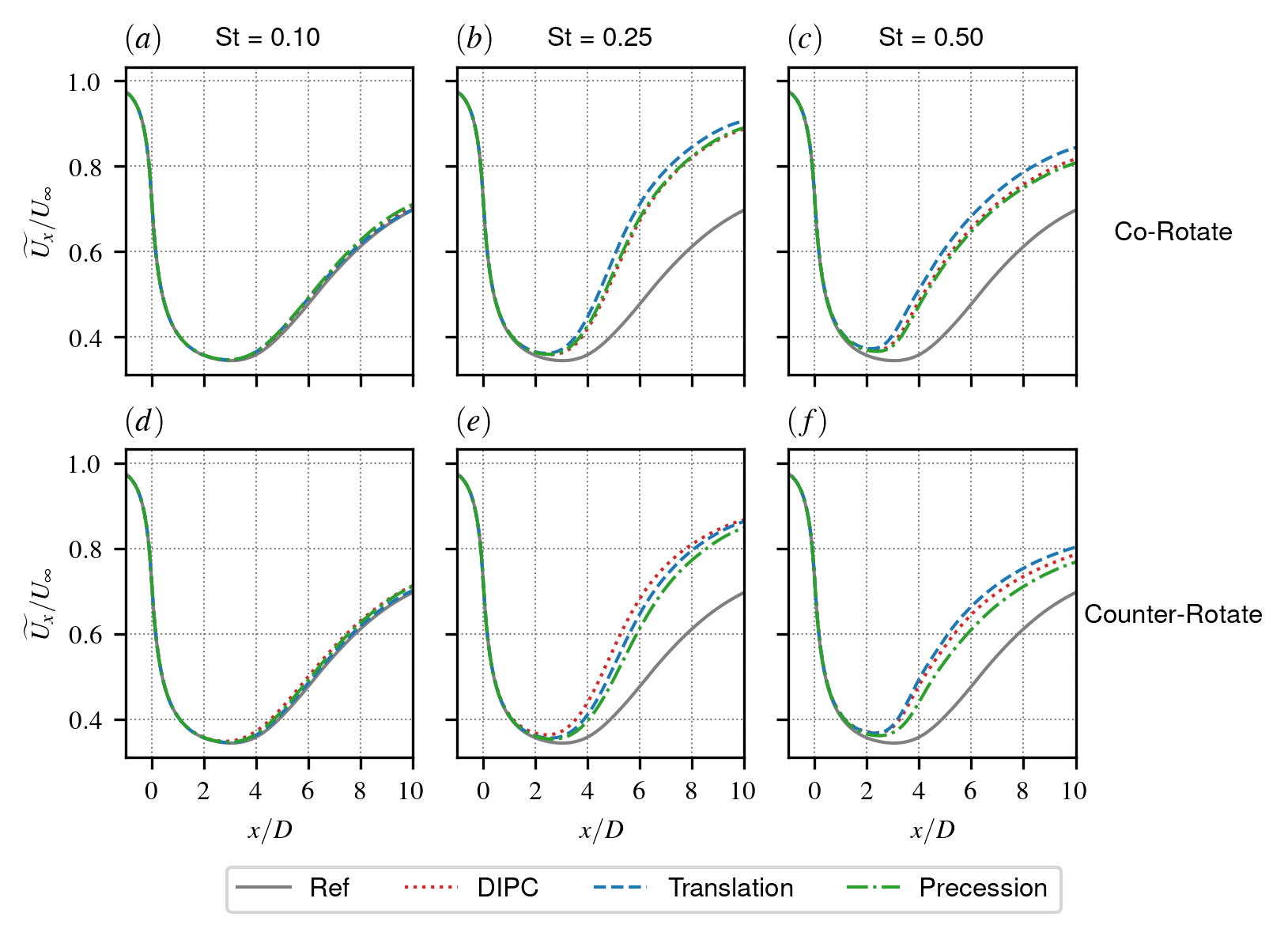}
    \caption{Comparison of time and rotor-area averaged wind speed $\widetilde{U_x}$ for different AWC approaches, forcing frequencies, and helix directions.}
    \label{fig:timeAndRotorAveragedWakeCurve}
\end{figure}

In this section, we investigate the recovery of the time-averaged wake velocity, which is the leading concern for wake mitigation controls  \citep{meyers2022wind}. 

Figure \ref{fig:timeAveragedWakeContour} presents time-averaged velocity field contours at the hub-height plane to reveal the influence of AWC approaches on wake recovery. The wake recovery exhibits strong frequency dependence for all three AWC techniques, with maximum recovery at $St = 0.25$, followed by progressively diminishing recovery effects at $St = 0.50$ and $St = 0.10$. For cases with strong wake recovery, the AWC fundamentally alters the spatial distribution of the wake velocity, characterised  by a broadened wake width and a dual-minima lateral profile. According to \cite{korb2023characteristics}, the non-Gaussian profile can be explained by the helical deflection of the instantaneous wake, which makes the peak of the probability density function of the instantaneous wake centre moves off the wake centreline. Such dual-peak distributions have also been observed in side-to-side meandering wakes \citep{li2022onset}.


Figure \ref{fig:timeAndRotorAveragedWakeCurve} shows the time- and rotor-area-averaged streamwise velocity ($\widetilde{U}_x$) as a function of downstream distance $(x)$ for various AWC approaches, compared to the baseline reference case without AWC. At $St=0.10$, all cases show minimal deviation from the reference, indicating negligible wake recovery. For $St=0.25$ and $St=0.50$, wake recovery is consistently enhanced regardless of AWC approach or helix direction. Specifically, AWC at $St=0.25$ accelerates wind speed more rapidly in the far wake, while $St=0.50$ achieves earlier recovery in the near wake $(x < 4D)$. This behaviour aligns with the strength of velocity fluctuations, which are more pronounced in the far wake at $St=0.25$ but are stronger in the near wake at $St=0.50$, as shown in figure \ref{fig:FFTVelocity_hubplane}. Additionally, the wake recovery enhancement varies with helix rotation direction, attributable to the non-zero rotational velocity in the baseline wake, as demonstrated by \cite{frederik2020helix}. Finally, the observed variations among AWC approaches in the far wake reflect the quantitative impact of different rotor actuations on the wake recovery enhancement.

\subsubsection{Reynolds stress \label{sec:RSS}}

\begin{figure}
    \centering
    \includegraphics[width=\textwidth]{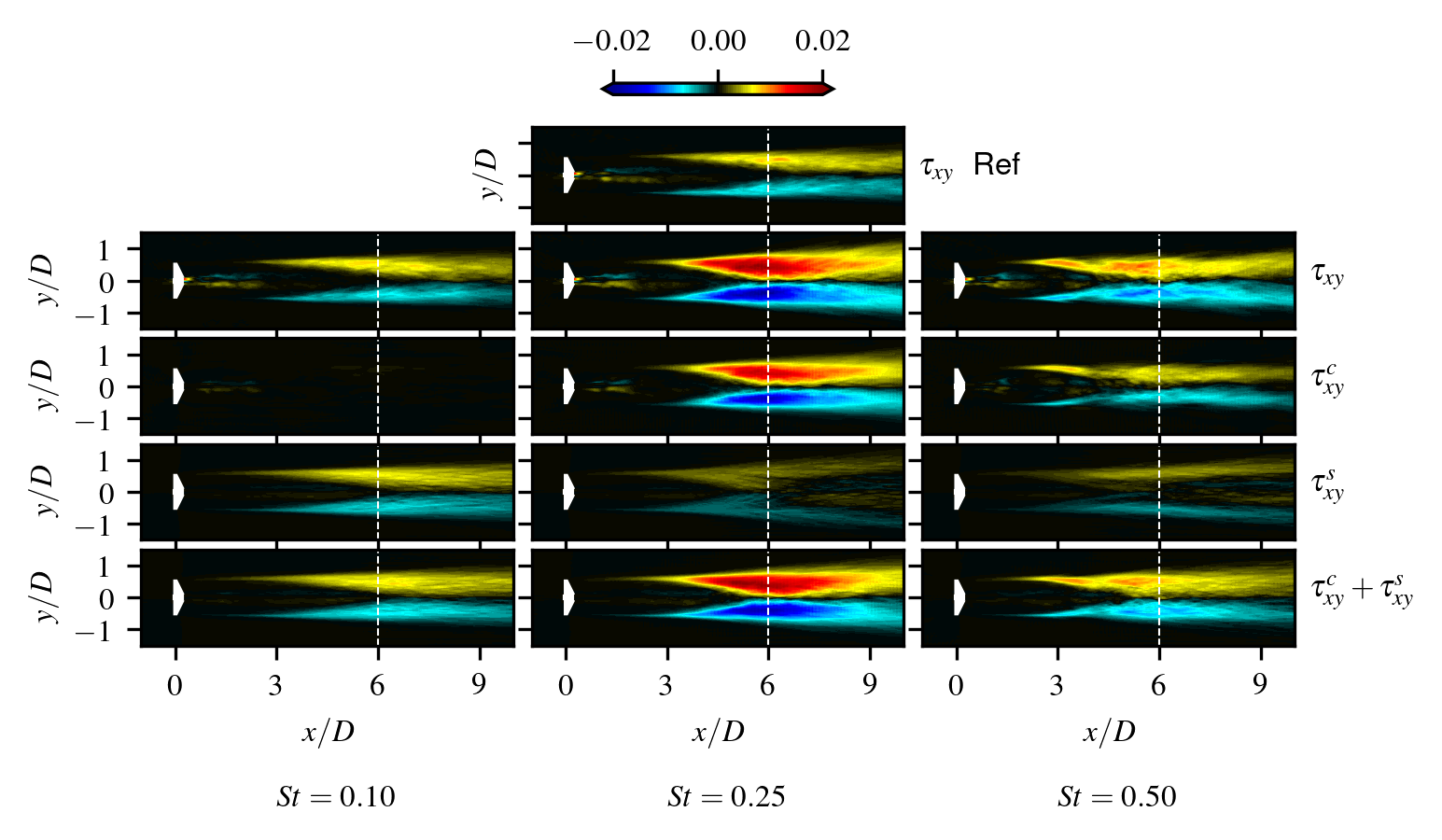}

    \caption{Contours of in-plane Reynolds shear stress on the hub-height plane for the reference case (first row) and cases with co-rotating DIPC. The second row shows the total shear stress ($\tau_{xy}$), the third row displays the coherent component ($\tau_{xy}^c$), the fourth row shows the incoherent component ($\tau_{xy}^s$), and the last row presents the sum of both components($\tau_{xy}^c + \tau_{xy}^s$). }
    \label{fig:ShearStressCompareContour}

    \centering
    \includegraphics[width=\linewidth]{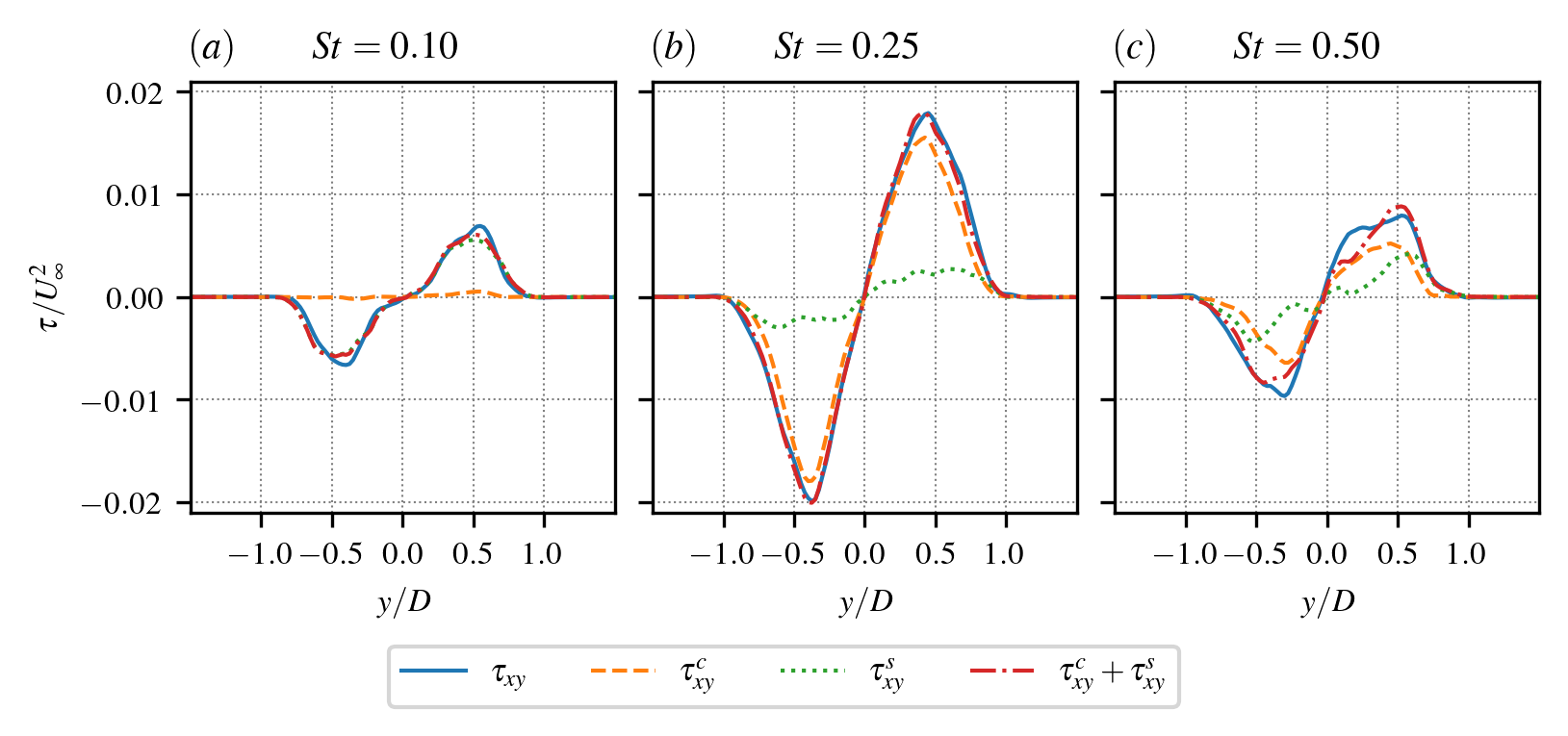}
    \caption{Same data with figure \ref{fig:ShearStressCompareContour} at $x=6D$ downstream of the rotor.}
    \label{fig:ShearStressCompareCurve}
\end{figure}

In  $\S$ \ref{sec:nonlinearModel}, we have followed the previous studies of shear flows to reconstruct the RS from contributions of the coherent structures ($\boldsymbol{\tau^c}$) and the incoherent small-scale turbulence ($\boldsymbol{\tau^s}$)  separately. This section aims to verify this decomposition strategy using LES results presented in figures \ref{fig:ShearStressCompareContour} and \ref{fig:ShearStressCompareCurve}.

Here, the coherent RS component is approximated by the coherent structures extracted from the LES result using Fourier transform $\boldsymbol{\tau^c}$ as follows:
\begin{equation}
    \tau^{c}_{ij,\text{LES}} =  -\frac{1}{2} \mathcal{R} \left(\hat{u}^{H}_{i,\text{LES}} \times \hat{u}_{j,\text{LES}} \right), \label{eqn:RSc}
\end{equation}
where $\hat{\{\cdot\}}_\text{LES}$ denotes the Fourier modes at the rotor forcing frequency ($\omega$).  The factor $\displaystyle \frac{1}{2}$ arises from the employed one-sided Fourier transform of a real time-domain signal, which is decomposed into the two-sided complex frequency-domain including both negative and positive frequencies ($\pm \omega$), see equation \eqref{eqn:spaceTimeDecomposition}. The incoherent small-scale RS component is approximated using an eddy viscosity and the meanflow strain rate, as follows 
\begin{equation}
\tau^s_{ij,\text{LES}} = \nu_\text{eff} \frac{\partial \overline{u}_{i,\text{LES}}}{\partial x_j},    
\end{equation}
where $\nu_\text{eff}$ is fitted from the baseline wake, according to equation \eqref{eqn:fitNut}. {\color{black}While all components of the Reynolds stress tensor are computed, we focus on the in-plane Reynolds shear stress due to its dominant role in momentum exchange and recovery of the time-averaged wake \citep{Gambuzza2023influence}.}

{\color{black}
In figure \ref{fig:ShearStressCompareContour}, we employ the in-plane Reynolds shear stress at the hub-height $(\tau_{xy})$ obtained from LES to analyse the contribution from coherent and incoherent small-scale parts.} The first row depicts {\color{black} $\tau_{xy}$} of the reference baseline case, i.e.,  {\color{black} $\tau_{xy}$} without AWC-induced coherent contribution. The subsequent rows show the  {\color{black} $\tau_{xy}$} subject to AWC. Here, DIPC is employed at various frequencies with co-rotating helix. The results plotted are the total  {\color{black} $\tau_{xy}$} obtained directly from LES ($\tau_{xy} = -\overline{u^{'}_x u^{'}_y}$), the coherent component ($\tau_{xy}^c$),  the incoherent component ($\tau_{xy}^s$), and the summation of both ($\tau_{xy}^c+\tau_{xy}^s$).  At $St=0.10$, the total $\tau_{xy}$ exhibits patterns remarkably similar to the reference case. The coherent component ($\tau_{xy}^c$) remains negligible throughout the wake region. In contrast, the incoherent component ($\tau_{xy}^s$) dominates their summation ($\tau^c_{xy} + \tau^s_{xy}$). This behaviour can be attributed to the weak coherent structures induced by the low-frequency forcing, as evidenced in figure \ref{fig:FFTVelocity_hubplane}. Additionally, the minimal modification of the mean wake at this frequency (figure \ref{fig:timeAveragedWakeContour}) also supports the strong similarity between the {\color{black}$\tau_{xy}^s$} and {\color{black}$\tau_{xy}$} of the reference case. At $St=0.25$, $\tau_{xy}$ increases in magnitude and expands in the far wake. {\color{black}$\tau_{xy}^c$} is greatly amplified and becomes the dominant part, due to the large-scale helical wake meandering. Conversely, {\color{black}$\tau_{xy}^s$} diminishes in magnitude, as the wake recovery reduces the mean velocity gradient. The sum of both components also closely approximates the directly computed $\tau_{xy}$. At $St=0.50$,  both extent and magnitude of $\tau_{xy}$ are notably reduced compared to $St=0.25$. While {\color{black}$\tau_{xy}^c$}  remains the primary contributor, its dominance diminishes. {\color{black}$\tau_{xy}^s$} shows a slight increase, attributed to the thinner wake shear layer and the increased velocity gradient. The superposition of both components also effectively approximates the directly computed $\tau_{xy}$.

To further confirm this decomposition strategy, figure \ref{fig:ShearStressCompareCurve} presents a quantitative comparison of the same data $6D$ downstream of the rotor. At $St=0.10$, the {\color{black}$\tau_{xy}$} is contributed almost exclusively by its incoherent component ($\tau_{xy}^s$). At $St=0.25$, the data demonstrates the significant enhancement of {\color{black}$\tau_{xy}$} due to increased {\color{black}$\tau_{xy}^c$}. At $St=0.50$, the overall magnitude of {\color{black}$\tau_{xy}$} decreases compared to $St=0.25$ and the relative contributions from {\color{black}$\tau_{xy}^c$} and {\color{black}$\tau_{xy}^s$} become comparable. Importantly, across the three characteristic scenarios, {\color{black}$\tau_{xy}^s + \tau_{xy}^c$} closely approximates the {\color{black}$\tau_{xy}$} directly computed from LES, and thus confirms the correctness of the present two-component RS modelling approach ($\S$ \ref{sec:RSSModel}). This validation extends consistently across all investigated AWC approaches, forcing frequencies and helix directions, though additional results are omitted for brevity.

\subsection{Model validation \label{sec:lesValidation}}

\begin{figure}
    \centering
    \includegraphics[width=\linewidth]{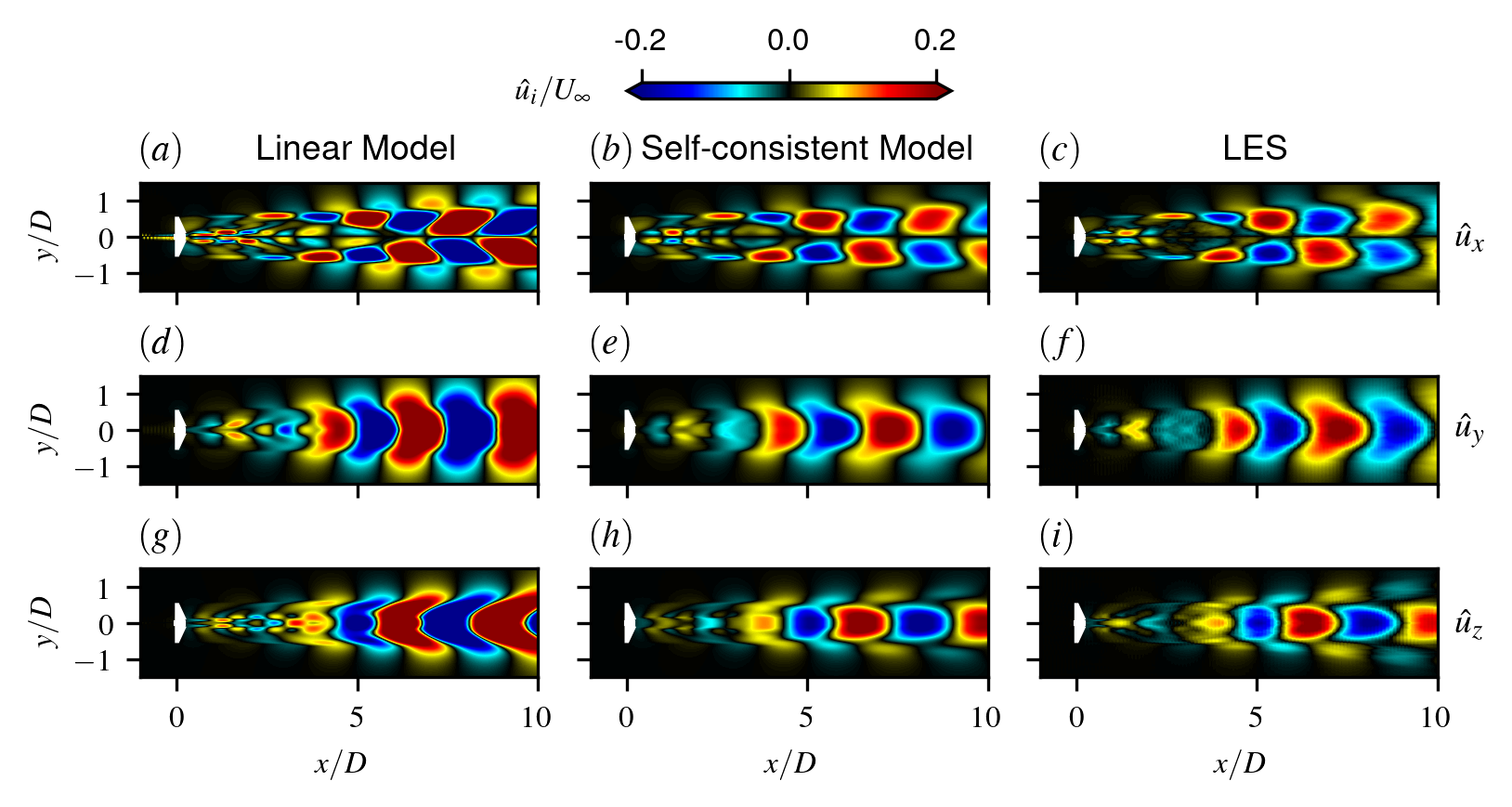}
    \caption{Coherent structures at hub height plane predicted by: (a)(d)(g) the linear resolvent model, (b)(e)(h)  the self-consistent model, and (c)(f)(i) large eddy simulations; (a)(b)(c) the streamwise velocity, (d)(e)(f) the spanwise velocity, and (g)(h)(i) the vertical velocity. AWC approach: DIPC at $St=0.25$ with co-rotate helix.}
   \label{fig:wakeFluctuationCompare_helix_CW025}

    \centering
    \includegraphics[width=0.75\linewidth]{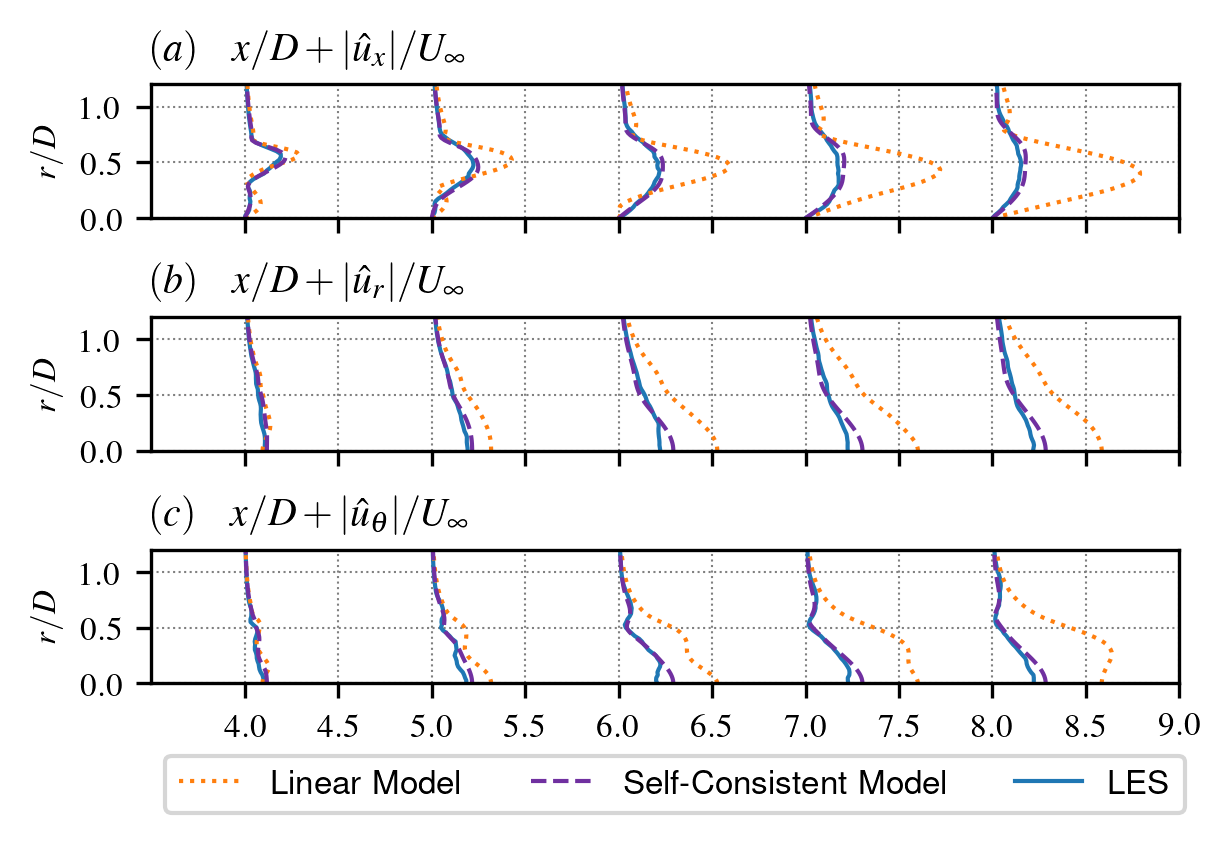}
    \caption{\color{black}Radial profiles of fluctuating velocity magnitudes of the same coherent structures shown in figure \ref{fig:wakeFluctuationCompare_helix_CW025} at downstream locations $x/D \in \{4,5,6,7,8\}$: (a) streamwise component, (b) radial component, (c) azimuthal component.     \label{fig:wakeFluctuationCompare_helix_CW025_curve}}
\end{figure}

This section validates the proposed self-consistent model through comparison with LES results for both coherent fluctuation and mean wake characteristics. Comparison to linear resolvent method is also conducted to demonstrate the improvement achieved by the proposed self-consistent model. Given the similarity among different control strategies, the flow field comparisons focus on the DIPC-forced wake. The section concludes with a quantitative analysis of AWC-induced wake recovery across various control strategies, forcing frequencies, and helix directions. 

{\color{black} For both linear and self-consistent models, a two-dimensional domain ($xOr$) extending $x \in [-2D,10.5D]$ and $r \in [0,3D]$ is employed, similar to that of LES. The grid interval is uniform in the streamwise direction with $\Delta x = D/20$. In the radial direction, the mesh is uniform in the wake with $\Delta r = D/40$, and is gradually stretched out for $r>1.2D$, resulting in $N_x \times N_r$ = $251 \times 71$ grid points. The mesh convergence is studied in Appendix \ref{sec:meshConv}}.

Figure~\ref{fig:wakeFluctuationCompare_helix_CW025}  presents a comparative analysis of coherent wake structures at the hub-height plane, comparing predictions from the self-consistent model, linear model, and LES, for DIPC-forced wake at $St=0.25$ with co-rotating helix. {\color{black} Figure~\ref{fig:wakeFluctuationCompare_helix_CW025_curve} compares coherent velocity oscillation magnitudes through radial profiles at multiple downstream positions within $4D \leq x \leq 8D$.} While both models capture the fundamental oscillatory patterns observed in LES, the self-consistent model achieves superior prediction accuracy in both amplitude and spatial distributions across all velocity components. The streamwise velocity components (figures~\ref{fig:wakeFluctuationCompare_helix_CW025} a-c) exhibit alternating positive and negative velocity regions, asymmetrically distributed about the wake centreline. Conversely, both spanwise (d-f) and vertical (g-i) velocity components display symmetric patterns about the centerline. 
The linear model exhibits known limitations \citep{li2024resolvent}. {\color{black}As shown in figures~\ref{fig:wakeFluctuationCompare_helix_CW025} (a, d, g), the linear model underestimates the wavelength of coherent structures, particularly in the far wake, since it does not account for the nonlinear effects leading to baseflow modification. Additionally, the stronger shear in the unmodified baseflow causes the linear model to overestimate velocity fluctuation magnitude, as clearly demonstrated in figure~\ref{fig:wakeFluctuationCompare_helix_CW025_curve}.}   The self-consistent model effectively addresses these limitations through several key improvements. {\color{black} Figure~\ref{fig:wakeFluctuationCompare_helix_CW025} (b, e, h) shows improved streamwise wavelength agreement with LES, reducing diamond-shaped deformation in streamwise velocity fluctuations compared to the linear model. Figure~\ref{fig:wakeFluctuationCompare_helix_CW025_curve} demonstrates superior prediction of fluctuating velocity magnitudes across all components, better capturing both qualitative and quantitative wake dynamics. Some discrepancies are observed in the contour plots between the self-consistent model and LES, particularly in the near wake, which may be attributed to the modelling simplification and the effect of nacelle in LES. }


\begin{figure}
    \centering
    \includegraphics[width=\linewidth]{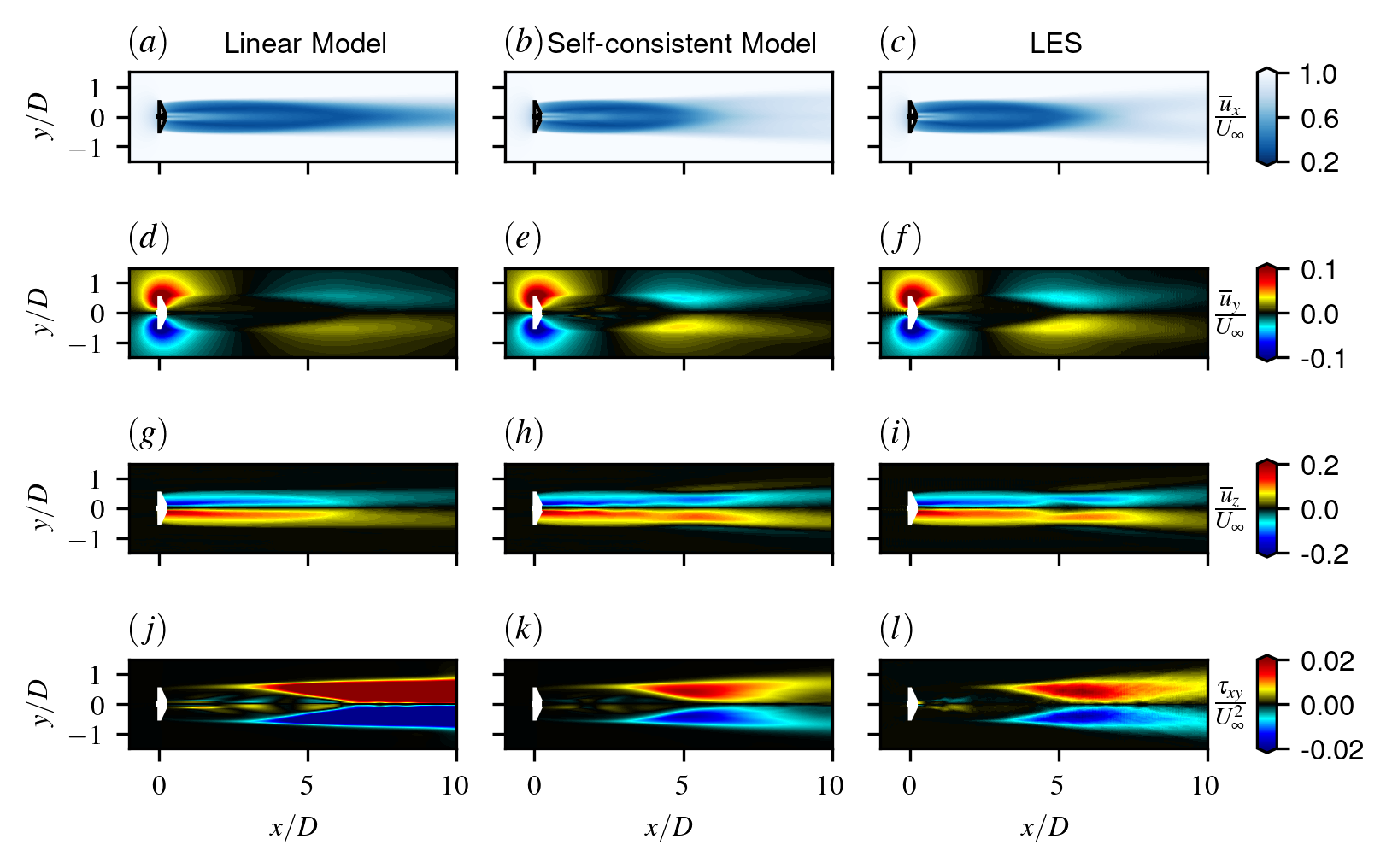}
    \caption{The time-averaged wake predicted by different methods: (a)(d)(g)(j) the linear resolvent model, (b)(e)(h)(k) the self-consistent model, and (c)(f)(i)(l) LES; (a)(b)(c) streamwise velocity, (d)(e)(f) spanwise velocity, and (g)(h)(i) vertical velocity, (j)(k)(l) $\tau_{xy}$. AWC approach: DIPC at $St=0.25$ with co-rotate helix.}
    \label{fig:wakeMeanFieldCompare_helix_025}


    \centering
    \includegraphics[width=0.75\linewidth]{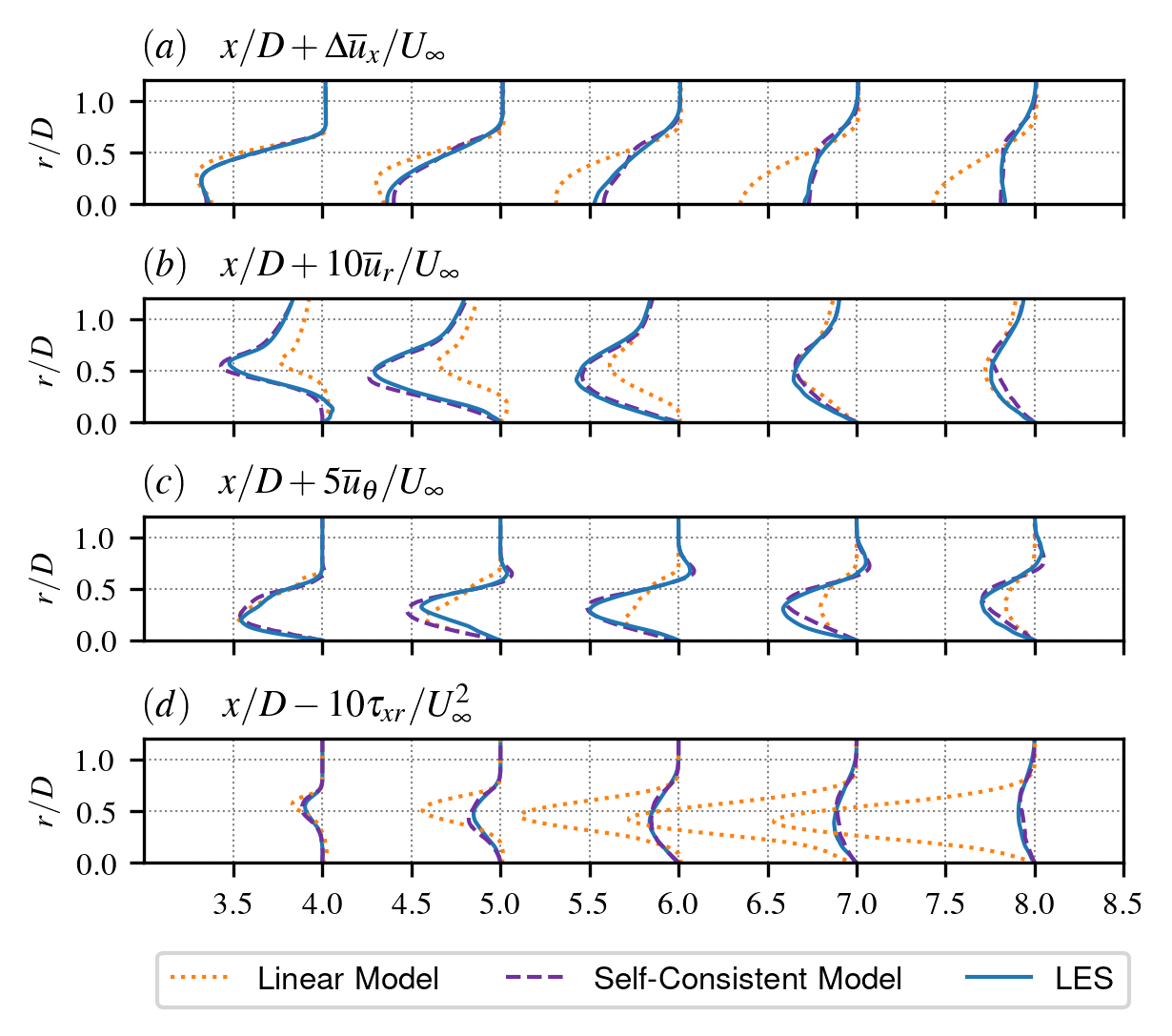}
    \caption{\color{black}Radial profiles of the same time-averaged wake characteristics shown in figure~\ref{fig:wakeMeanFieldCompare_helix_025} at downstream locations $x/D \in \{4,5,6,7,8\}$. }
    \label{fig:wakeMeanFieldCompare_helix_025_curve}


\end{figure}

Figure~\ref{fig:wakeMeanFieldCompare_helix_025}  presents a comprehensive comparison of the time-averaged wake characteristics at the hub-height. The comparison encompasses the streamwise (a-c), spanwise (d-f), and vertical (g-i) velocity components, along with the in-plane RS distribution (j-l). {\color{black} Figure~\ref{fig:wakeMeanFieldCompare_helix_025_curve} plots the same data using radial profiles at downstream locations $4D \le x \le 8D$.}  For the linear model, the velocity field is equal to the reference baseline wake, since the linear model does not include any mean wake modification, the in-plane Reynolds shear stresses ($\tau_{xy}$ and $\tau_{xr}$) for the linear model are post-processed using equations (\ref{eqn:RSs} - \ref{eqn:RSs2}) but not employed to correct the mean wake. As seen, the self-consistent model successfully captures the modification made by AWC on the time-averaged field across all velocity components, whereas the linear model is inherently incapable. {\color{black}For the streamwise velocity, the self-consistent model accurately reproduces the apparent expansion and increased wind speed in the far wake, showing excellent agreement with LES results (figures~\ref{fig:wakeMeanFieldCompare_helix_025} b, c, and figure~\ref{fig:wakeMeanFieldCompare_helix_025_curve} a). For the spanwise component, an increased converging velocity from the freestream towards the wake centre is observed around $x=5D$ in both the prediction of the self-consistent model and LES (figures~\ref{fig:wakeMeanFieldCompare_helix_025} e, f,  and figure~\ref{fig:wakeMeanFieldCompare_helix_025_curve} b), indicating that the enhanced flow entrainment to the wake region is well captured by the proposed model. Moreover, an improved agreement of the out-plane velocity ($\overline{u}_z$ in Cartesian coordinates and $\overline{u}_\theta$ in the cylindrical coordinates) is found between the self-consistent model and LES at $x>5D$, indicating that the proposed model captures the increased rotational velocity due to the co-rotating helix (figures~\ref{fig:wakeMeanFieldCompare_helix_025} h, i, and figure~\ref{fig:wakeMeanFieldCompare_helix_025_curve} c). Lastly, $\tau_{xy}$ ($\tau_{xr}$) predicted by the self-consistent model achieves improved agreement with LES in both magnitude and spatial distribution (figures~\ref{fig:wakeMeanFieldCompare_helix_025} k, l, and figure~\ref{fig:wakeMeanFieldCompare_helix_025_curve} d), explaining the model's success in time-averaged wake prediction, since it is the key factor driving wake recovery in the mean momentum equation~\eqref{eqn:meanMom}. }

\begin{figure}
    \centering
    \includegraphics[width=\linewidth]{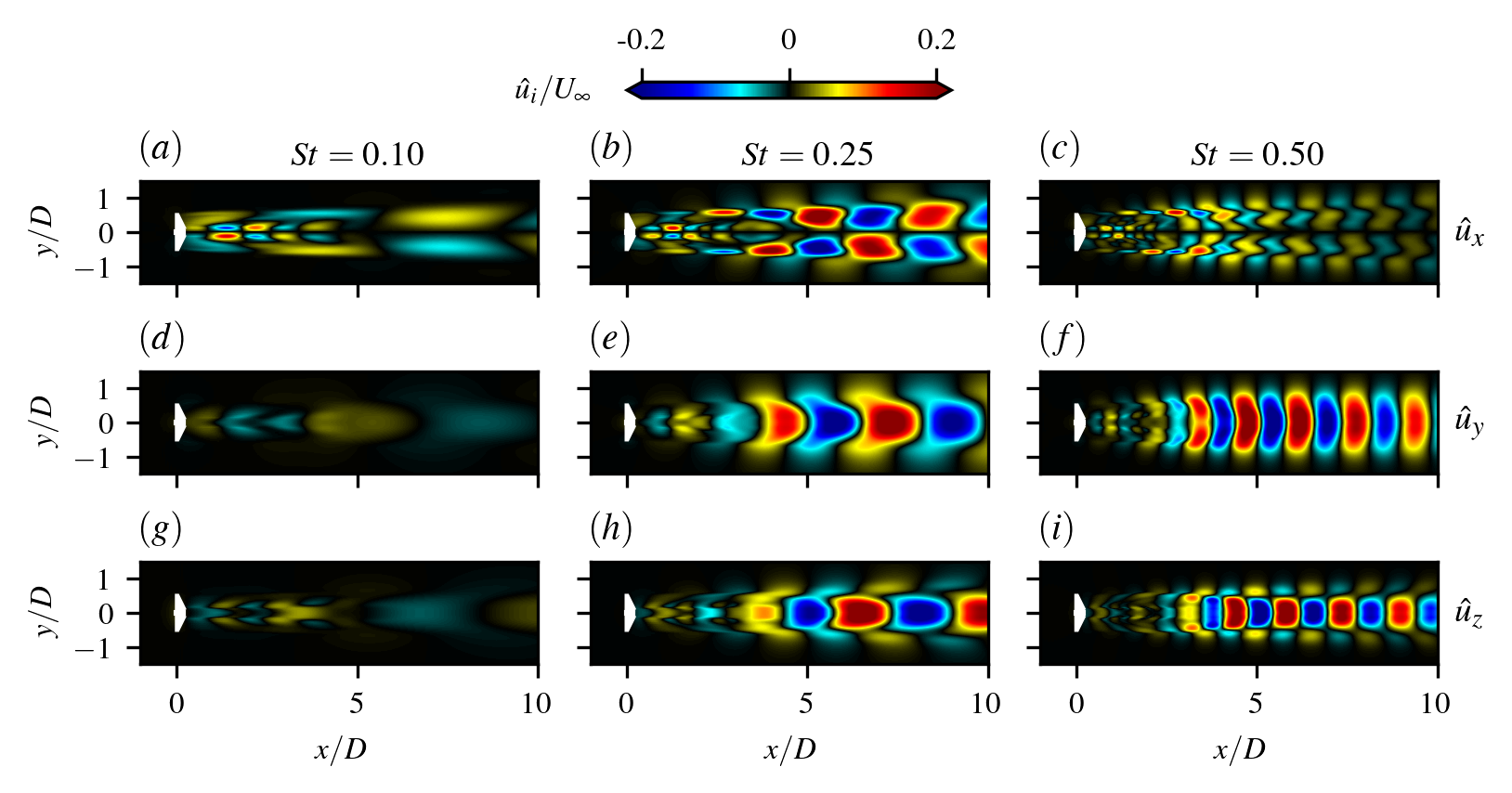}
    \caption{Wake response predicted by the self-consistent model at different frequencies: $St = 0.10$ (a, d, g), 0.25 (b, e, h) and 0.50 (c, f, i). AWC approach: DIPC with co-rotating helix.}
    \label{fig:resolventResponse_helix_nonlinear}
    \includegraphics[width=0.75\linewidth]{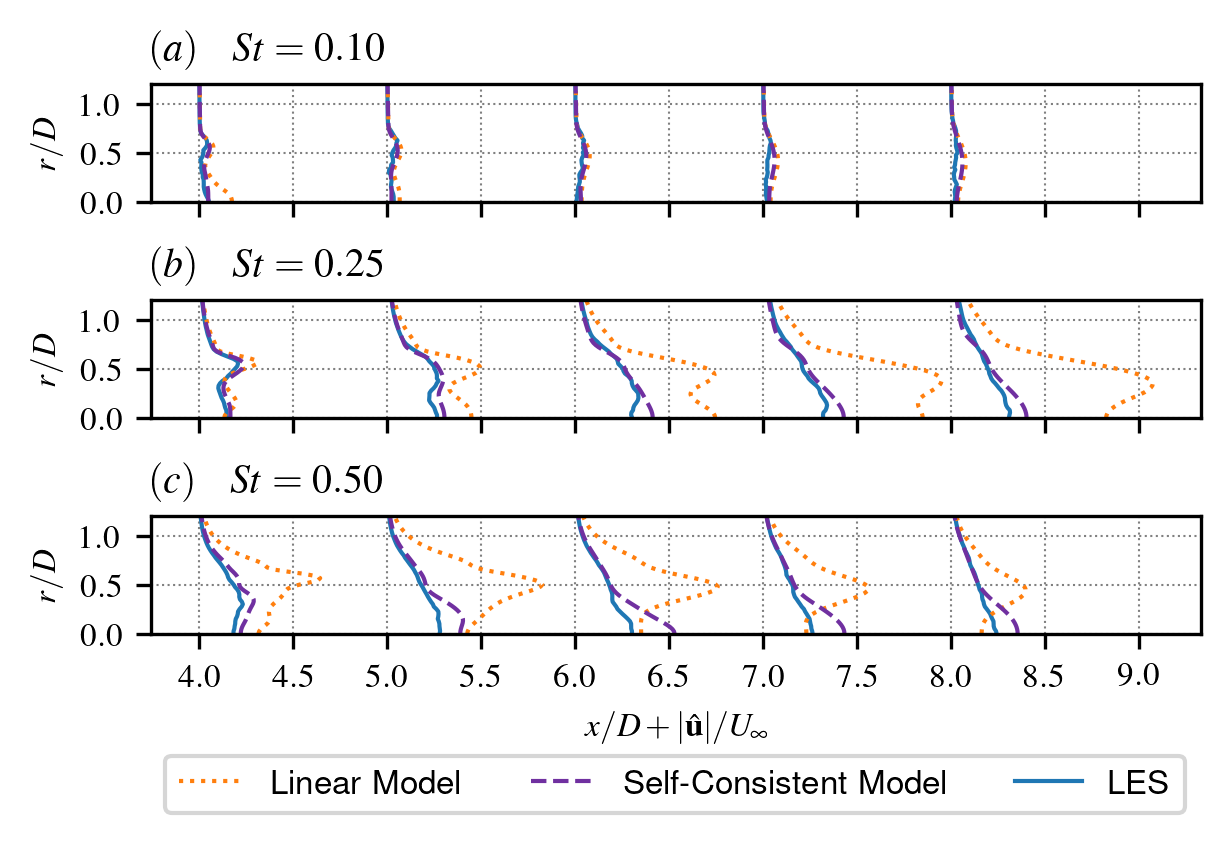}
    \caption{\color{black}Comparison of radial profiles of the total fluctuating velocity magnitude $|\hat{\mathbf{u}}| = \sqrt{\hat{u}^2_x + \hat{u}^2_r +\hat{u}^2_\theta}$ predicted by linear model (figure \ref{fig:resolventResponse_helix}), self-consistent model (figure~\ref{fig:resolventResponse_helix_nonlinear}), and LES (figure \ref{fig:FFTVelocity_hubplane})  at downstream locations $x/D \in \{4,5,6,7,8\}$: (a)~$St = 0.10$, (b)~$St = 0.25$, (c)~$St = 0.50$.  AWC approach: DIPC with co-rotating helix.}
    \label{fig:modelCompareAtDifferentSt}
    
\end{figure}

Figure~\ref{fig:resolventResponse_helix_nonlinear} presents  wake responses predicted by the self-consistent model at $St = 0.10$ (a, d, g), $St = 0.25$ (b, e, h), and $St = 0.50$ (c, f, i). {\color{black}Figure~\ref{fig:modelCompareAtDifferentSt}
compares the magnitude of the coherent velocity fluctuation predicted by the  
linear model (figure \ref{fig:resolventResponse_helix}), the self-consistent model  (figure  \ref{fig:resolventResponse_helix_nonlinear}), and the LES (figure \ref{fig:FFTVelocity_hubplane}) by plotting radial profiles of the fluctuating velocity magnitude ($|\hat{\mathbf{u}}| = \sqrt{\hat{u}^2_x + \hat{u}^2_r +\hat{u}^2_\theta}$) at downstream locations in the far wake ($4D \le x \le 8D$). At $St=0.10$, the difference between linear model, self-consistent model, and LES is minimal due to weak AWC-induced wake response and minimal nonlinearity. For the case with $St=0.25$, where the large-scale meandering is the most prominent, good agreement between the self-consistent model and LES is found. For $St = 0.50$, although discrepancies between the self-consistent model and LES exist, the overestimation of fluctuating velocities is alleviated compared to the linear model.}

\begin{figure}
    \centering
    \includegraphics[width=\linewidth]{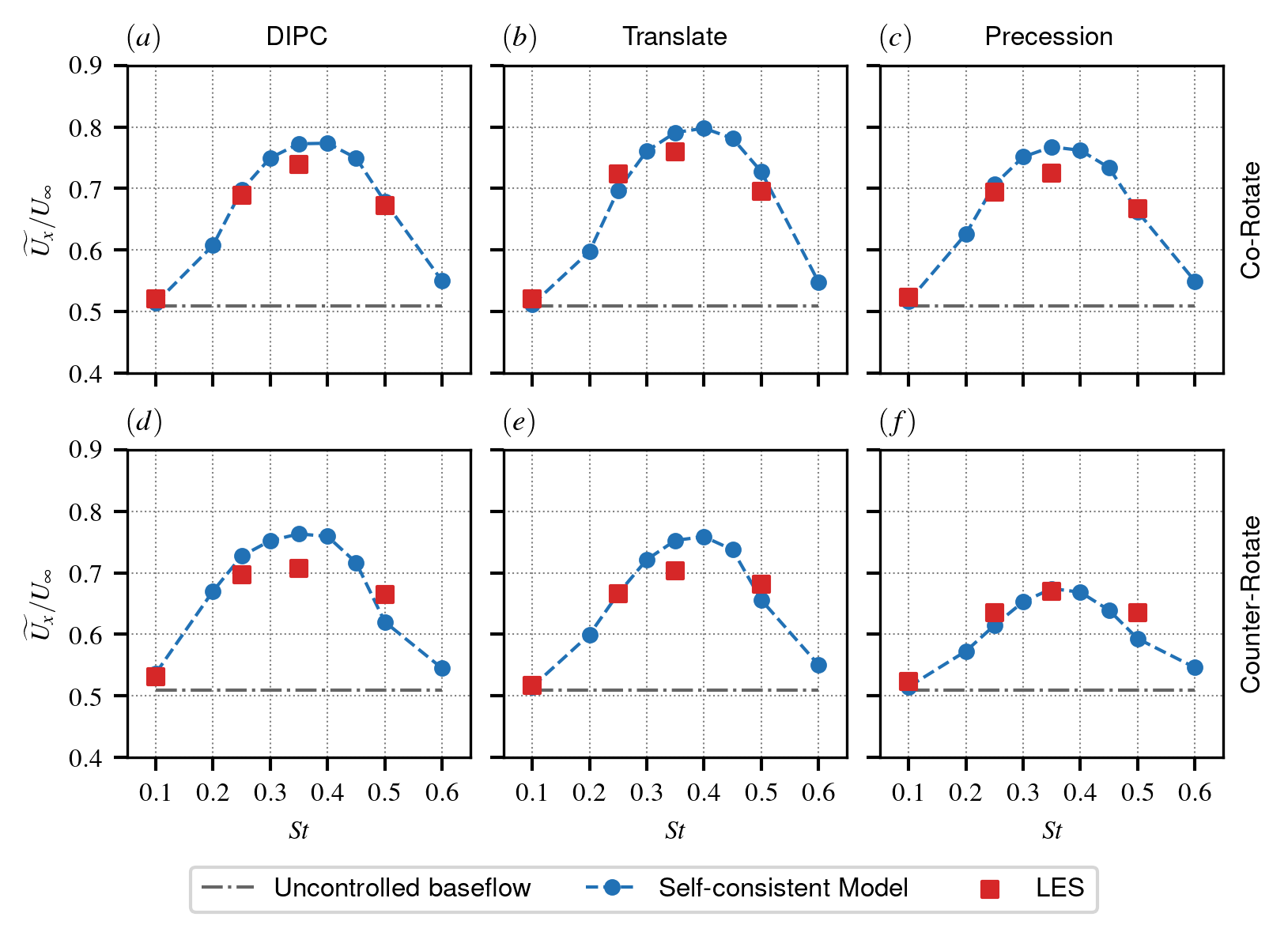}
    \caption{The time- and rotor-area-averaged streamwise velocity predicted by the self-consistent model and large eddy simulation for all considered AWC approaches: DIPC (a)(d), translate (b)(e), precession (c)(f). The helix direction is co-rotate and counter-rotate for the top and bottom rows, respectively. {\color{black}The uncontrolled baseflow is provided for reference.} The results are plotted at $x=6D$.}
    \label{fig:compareWakeRecoveryAll}


\end{figure} 

Finally, the predictive capabilities on the enhanced wake recovery of the self-consistent model are quantitatively  assessed. Here, the self-consistent model is employed to predict wake velocity over a broad range of forcing frequencies ($0.1 \le St \le 0.6$), for all three forcing strategies, and with both helix directions. Figure \ref{fig:compareWakeRecoveryAll} presents the time- and rotor-area-averaged streamwise velocity ($\widetilde{U}_x$) at $6D$ downstream the rotor, comparing predictions from the linear resolvent model (without meanflow correction), the self-consistent model, and LES. Compared to the baseline state, the self-consistent wake model successfully predicts the enhanced wake recovery following the same frequency dependency trend showed by LES results: the highest recovery occurs around $St \approx 0.35$ and diminishes when the frequency moves toward both extremities for all considered AWC approaches. Moreover, quantitative agreement is found between the self-consistent model and LES for all the considered cases, with the largest relative discrepancy less than 8\% (the maximum error occurs for the counter-rotating DIPC case at $St=0.35$).  The discrepancies between the self-consistent model and LES may be associated with the simplifications in the modelling, and shows the room for further improvements. Even in this least favourable scenario, the velocity prediction accuracy is still remarkably improved against the linear model, which excludes the meanflow modification. For the linear model, the time-averaged wake is always equal to the uncontrolled baseline wake, which shows a 
28.0\% underestimation of the mean wake velocity.  Overall, the self-consistent model demonstrates a reliable and accurate mean velocity prediction across diverse control strategies, demonstrating its potential as a computationally efficient alternative to LES for AWC designs.

{
\color{black}
\section{Discussion} \label{sec:discussion}  

The model validation presented above demonstrates promising capabilities for predicting enhanced wake recovery and fluctuating velocities under AWC. Furthermore, the model's gain in computational efficiency is substantial, achieving convergence within ten minutes on a desktop computer compared to LES simulations requiring $5 \times 10^3$ CPU hours. This efficiency enables exploration of extensive parameter spaces—such as the 54 cases shown in figure \ref{fig:compareWakeRecoveryAll}—overnight. The computational  cost can be further reduced by employing coarser resolution as shown in Appendix \ref{sec:meshConv}. This combination of accuracy and computational efficiency establishes the model as a practical tool for AWC design and optimization.

The predictive capability and efficiency of the model are achieved through several simplifications throughout its development: the axisymmetric baseflow, uniform non-turbulent inflow, and frequency lock-in between AWC and wake response. Here, we discuss key considerations regarding the physical mechanisms captured by the model, its applicability to real-world conditions, and potential pathways for future refinement. 

First, the unified Eulerian perspective employed by the present model for both wake meandering and the mean wake recovery is found to well capture the flow physics. Commonly, wake meandering is modelled in the Lagrangian perspective, where the instantaneous wake deficit is treated as a passive scalar oscillating laterally in cross-stream directions, while the turbulent mixing between the instantaneous wake and freestream by small-scale turbulence is modelled in the Eulerian framework \citep{larsen2008wake}.  In the present model, the meandering is also modelled in the Eulerian framework as coherent velocity fluctuations, such that the instantaneous meandering motion of the wake deficit is represented by fluctuating streamwise velocity asymmetrically distributed along the time-averaged wake centreline as shown in figure \ref{fig:resolventResponse_helix_nonlinear} (b), and its modification effect on the time-averaged wake is computed via Reynolds stress. The turbulent mixing contributed by small-scale incoherent turbulence is modelled through the eddy viscosity $\nu_\text{eff}$, applied to both fluctuating velocity and meanflow. The present model properly captures the combined contribution of both coherent and incoherent parts as shown in figure \ref{fig:wakeMeanFieldCompare_helix_025_curve} (d). When ambient turbulence is included, both contributions may be influenced. We speculate that the present approach remains valid when the inflow turbulence is rather weak, i.e., close to the present non-turbulent case, which coincides with the weak ambient turbulence condition where AWCs are considered more effective \citep{frederik2020helix,li2022onset}. The challenges and opportunities for extending the model to realistic atmospheric conditions will be discussed in the third point of this section.

Second, our current model demonstrates that AWC-enhanced wake recovery is dominated by far-wake shear layer Kelvin-Helmholtz instability forced by rotor-level forcing. At the present relatively low forcing frequencies ($0.1 \le St \le 0.6$), the blade-level dynamics and tip-vortices in the near wake can be sufficiently simplified. While our model focuses on these lower frequencies, higher frequency perturbations associated with blade-level forcing and tip-vortex dynamics may also play important roles in wake evolution, as highlighted by recent work from \citet{biswas2024energy}. Their work demonstrated that velocity fluctuations at rotor frequency and its superharmonics appear in the near wake. For the present rotor with a tip speed ratio $\lambda = \Omega R/U_\infty = 9$, the rotor frequency corresponds to approximately $St_\text{rotor} \approx 2.9$, computed as:  
\begin{equation}  
    St_\text{rotor} = \frac{f_\text{rotor}D}{U_\infty} = \frac{\Omega \cdot 2 \cdot R}{2\pi U_\infty} = \frac{\lambda}{\pi} \approx 2.9  
\end{equation}  
Due to the significant frequency separation between the tip-vortices and the AWC-induced wake dynamics, their direct coupling is not explicitly considered in our model. Instead, the effects of these smaller-scale flow structures on large-scale coherent velocity fluctuations and mean wake are modelled through the effective viscosity $\nu_\text{eff}$.  The generally good agreement between the model's predictions and the tip-vortex-inclusive LES results not only validates the present modelling but also reveals that the shear-layer's Kelvin-Helmholtz instability is the dominant contributor to the enhanced wake recovery at the present low forcing frequency range of AWCs. 

Third, one limitation of the present study is that it employs uniform inflow. Extending it to real-world scenarios, where background turbulence, shear and veer exist, requires further in-depth study on the interaction of wake flow and atmospheric boundary layer (ABL).  ABL conditions typically involve shear, veer and varying levels of turbulence intensity, which can alter the development and stability characteristics of wind turbine wakes. Previous studies have shown that despite ambient turbulence, AWC still produces significant wake enhancement, though with varying effectiveness. For instance, \citet{frederik2020helix} demonstrated that although ambient turbulence affects wake dynamics, there remains a remarkable enhancement of wake recovery under AWC. Similarly, \citet{li2022onset} found that motion-induced wake recovery enhancement is stronger under low ambient turbulence conditions, but decreases with increasing ambient turbulence level. Not only in the context of AWC,  the interaction between turbulent wakes and ABL turbulence continues to present significant challenges for both fundamental fluid mechanics research and wind energy applications, as noted by \citet{stevens2017flow}.  Ongoing research continues to investigate the flow mechanisms of wake-ABL interaction. \citet{kleusberg2019tip} investigated shear effects on tip-vortex behaviour, while others have examined how ambient turbulence intensity \citep{Gambuzza2023influence} and length scales \citep{li2024impacts} alter wake development. From the modelling point of view, incorporating inflow shear and turbulence would require modifications to both the baseline flow and the resolvent operator of the present model. Three dimensional resolvent analysis will be required to account for the effect of the non-axisymmetry induced by non-uniform inflow, which requires advanced numerical solution techniques to reduce the prohibitive computational cost issue \citep{towne2022efficient}. These considerations, from both fluid mechanism and computational methods aspects, represent important directions for future research to understand the wake dynamics in realistic atmospheric conditions and to enhance the predictive capability of wake models. 

Lastly, the current implementation of our model relies on time-averaged LES data to establish the baseline wake flow, which serves as one of the two inputs of the wake model. Here, we emphasize that only one baseline LES is needed for all the 54 analyses presented in figure \ref{fig:compareWakeRecoveryAll}, i.e., for different AWCs, different control frequencies, and different helix directions. However, we acknowledge that dependence on high-fidelity simulations limits the model's stand-alone applicability for rapid engineering assessments. In future studies, several pathways for improvement can be considered: First, simplified analytical models for the baseline wake, such as those based on Gaussian profiles or self-similar solutions \citep{bastankhah2014new}, could be integrated with the resolvent framework. Second, the model could be coupled with faster computational approaches, such as Reynolds-Averaged Navier-Stokes (RANS) simulations or their variants \citep{ainslie1988calculating}. Third, data-driven machine-learning-based models could provide a compromise between computational efficiency and physical accuracy \citep{li2024physics}. These approaches, while promising for eliminating dependence on LES data, require systematic validation against experimental or high-fidelity numerical data to establish their predictive capabilities and thus require further investigations.
}

\section{Summary and conclusion \label{sec:conclusion}}

This study introduces a computationally efficient end-to-end predictive model for wind turbine wake under active wake control (AWC). Unlike the existing models based on linear stability or resolvent analysis, our framework incorporates the nonlinearly coupled interaction between wake fluctuations and meanflow modification, enabling self-consistent predictions of both wake meandering motion and enhanced recovery under various AWC.

The proposed modelling framework integrates four physics-based elementary submodels, i.e., a rotor forcing model, a resolvent model, a Reynolds stress model, and a meanflow model. The inputs of the model are: a user-defined AWC to be evaluated and a non-actively-controlled baseline wake. In the proposed framework,  the rotor forcing, coherent wake response, Reynolds stress, and time-averaged wake recovery enhancement are predicted sequentially by each submodel. Once the time-averaged wake recovery is predicted at one step, it is fed back as a baseflow for solving the coherent wake response in the next iteration, so that the mutual influence of the fluctuation and mean of the wake is considered, and a self-consistent prediction of the AWC-induced coherent fluctuation and the enhanced wake recovery can be achieved upon convergence. 

To the best of the authors' knowledge, this study presents the first fast running wake model capable to predict AWC-specific wake evolution in a quantitatively correct manner. The model is able to distinguish the difference made by different AWC techniques, and demonstrates superior predictive capabilities for both coherent structures and time-averaged wake characteristics. The framework successfully resolves the fluctuation overestimation issue of the linear resolvent method and accurately captures the time-averaged wake modification. These time-averaged wake modifications include enhanced streamwise velocity recovery, augmented centerline-converging spanwise velocity, and modified rotational velocity. Extensive validation against LES across various AWC approaches, forcing frequencies, and helix rotating directions confirms the model's robust predictive capabilities. The model maintains high accuracy across a broad frequency range ($0.1 \le St \le 0.6$) in predicting frequency-dependent wake recovery enhancement. This level of accuracy represents a significant advancement over linear models, which are inherently unable to capture wake recovery enhancement. {\color{black}The success of the present modelling approach also demonstrates the dominant role of shear layer instability in enhancing wake recovery within this low-frequency regime, compared to other wake flow structures such as the near-wake tip-vortex, which is not explicitly accounted for in the model.}

Future development of this framework may extend it to more complex baseflow configurations. Recent advances in efficient global resolvent analysis \citep{martini2021efficient,Ribeiro2020,towne2022efficient} provide potential pathways for expanding the current axisymmetric formulation to three-dimensional wake dynamics. Such extensions could enable the analysis of asymmetric phenomena, including ground effects, inflow shear/veers, thermal stability, and wake deflection strategies that produce non-axisymmetric meanflow. Another interesting direction would be extending the present framework further to address realistic atmospheric conditions, building upon recent investigations of turbulence effects \citep{Gambuzza2023influence, wu2023composition, li2024impacts} and inflow unsteadiness \citep{gupta2019low,Wei_2024}. {\color{black} Extending the forcing to include multiple frequencies and to combine multiple AWC types would also open a broad domain for investigating their individual responses and potential interactions.}

{
\color{black}
\backsection[Acknowledgements]{The authors gratefully acknowledge the anonymous reviewers for their insightful comments on the numerical implementation and interpretability of the model.}
}

\backsection[Funding]{
This work was supported by the Basic Science Center Program for ``Multiscale Problems in Nonlinear Mechanics" of the National Natural Science Foundation of China (No.12588201, No. 11988102), the Strategic Priority Research Program of Chinese Academy of Sciences (No. XDB0620102), and NSFC (No. 12172360, No.12202453). The LES in this study were carried out on the ORISE Supercomputer.
}

\backsection[Declaration of interests]{The authors report no conflict of interest.}

\backsection[Data availability statement]{The data that support the findings of this study are available from the corresponding author upon reasonable request.}

\appendix
{\color{black}
\section{Solution Convergence}\label{appA}

\begin{figure}
    \centering

    \includegraphics[width=0.85\linewidth]{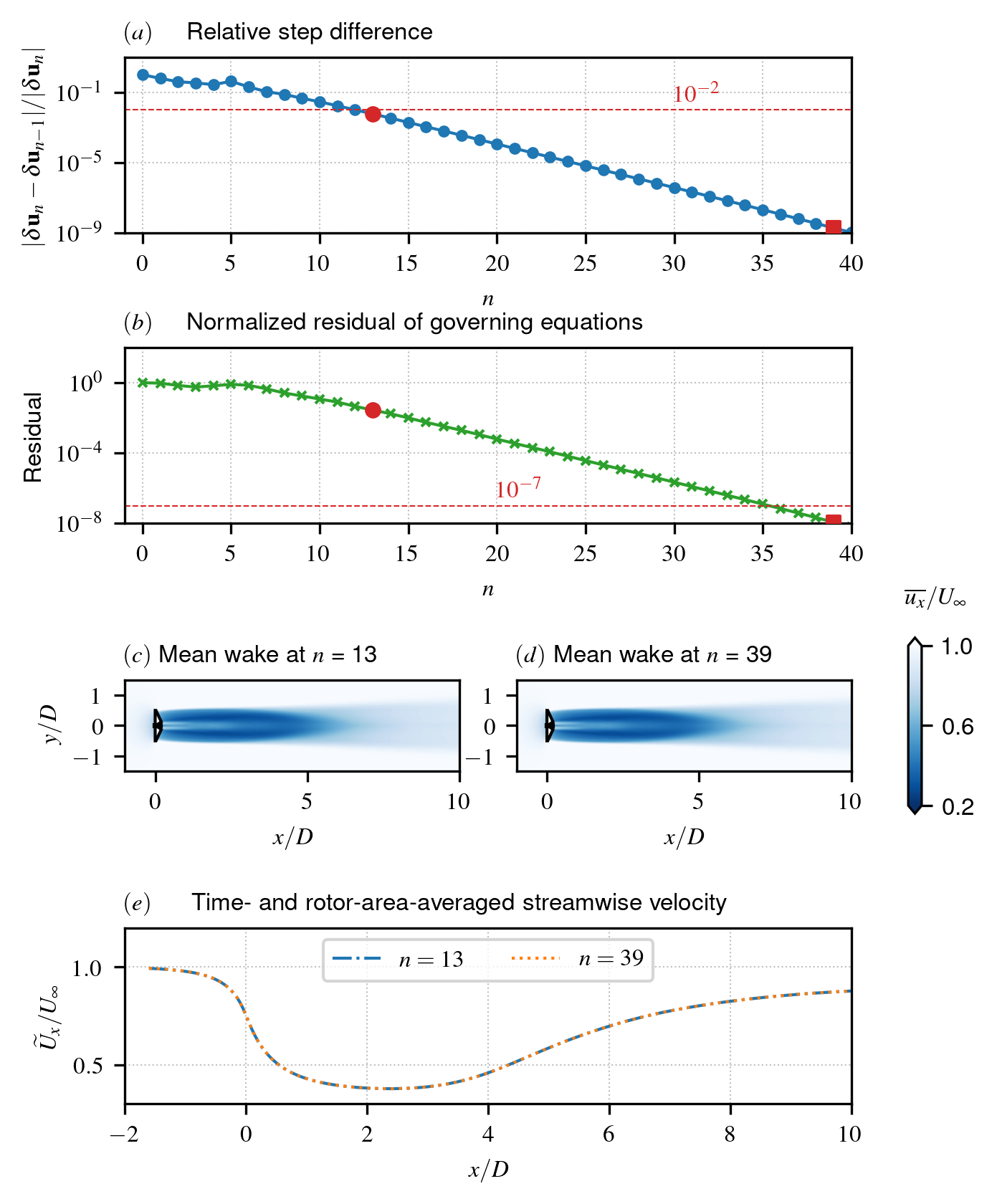}
    \caption{Effect of stopping criteria on self-consistent model solutions : (a) relative change in meanflow modification between successive iterations; (b) normalized residual of the governing equations; (c) and (d) mean streamwise velocity at iterations $n=13$ and $n=39$, respectively; (e) streamwise variation of time- and rotor-area-averaged streamwise velocity at iterations $n=13$ and $n=39$. Test case is the same as figure \ref{fig:IterativeResults} }
    \label{fig:stepConvergence}
\end{figure}

\subsection{Stopping criteria of nonlinear iteration \label{sec:stoppingcriteria}} 

The model uses a fixed-point iteration scheme to solve the coupled nonlinear system of coherent velocity fluctuation and meanflow correction. Solution convergence is monitored by the relative step difference, as detailed in Section \ref{sec:iterativescheme}. However, this approach may stagnate prematurely, yielding solutions that fail to satisfy the governing equations. To examine this issue, we extend the number of iterations and simultaneously monitor both the relative step difference in meanflow modifications and the equation residuals. The residual vector is calculated as the difference between the right and left sides of the equation system (\ref{eqn:coherentStructuresMom}- \ref{eqn:coherentStructuresIncom} and \ref{eqn:meanCorrMom}-\ref{eqn:meanCorrIncom}). It is quantified using the L2 norm via Python's \texttt{numpy.linalg.norm()} function and is then normalized by the L2 norm of the source terms of the same equations. 

Figure \ref{fig:stepConvergence} illustrates the convergence behaviour of our nonlinear iteration scheme. As shown in figure \ref{fig:stepConvergence} (a), the relative step difference decreases monotonically beyond the first several iterations, reaching approximately $10^{-9}$ at $n=39$. The normalized residual of the governing equations, presented in figure \ref{fig:stepConvergence} (b), exhibits a similar reduction trend. It is approximately $2\times 10^{-2}$ at $n = 13$ and attains a magnitude of about $10^{-8}$ at $n = 39$. This residual could be further reduced with additional iterations but would ultimately be limited by the precision of the linear solver or machine precision ($\sim10^{-7}$ and $\sim10^{-16}$ for single and double precision floating-point, respectively). Despite the decrease in equation residual with increasing number of iterations, figures \ref{fig:stepConvergence} (c-d) reveal that the mean wake prediction has effectively stabilized at $n=13$, compared with that at $n=39$. This observation is further confirmed in figure \ref{fig:stepConvergence} (e), where the time- and rotor-area-averaged streamwise velocity profiles at these iterations are virtually indistinguishable, with less than 0.1\% relative difference at $x = 6D$ for the present case. Considering that the computational cost is approximately proportional to the number of iterations, we have employed a relative step difference criterion of $10^{-2}$ (reached at $n=13$ for this case) as the stopping criterion in this paper, to balance computational efficiency with solution accuracy.

\subsection{Mesh convergence and computational cost \label{sec:meshConv}}

This appendix examines the influence of grid resolution on the self-consistent model solutions. The finest mesh, with a streamwise grid interval of $\Delta x = D/20$, corresponds to that used in the LES simulations and has been employed by the model throughout the main body of this paper. To assess convergence, three additional grids with progressively coarser resolutions were created, with the coarsest having a streamwise grid interval of $\Delta x = D/12$. The grid spacing in the radial direction was scaled proportionally to the streamwise direction for all cases. The DIPC approach with a co-rotating helix at a Strouhal number of $St = 0.25$ was selected as the test case.  

Figure \ref{fig:meshConvergence} compares the results obtained across the four different grid resolutions, showing: (a) the fluctuating velocity magnitude $|\hat{\mathbf{u}}|/U_\infty$ and (b) the time-averaged wake deficit $\Delta \overline{u}_x$. As evident from the figure, the different grid resolutions produce remarkably similar results, with particularly close agreement when $\Delta x$ is finer than $D/14$. A minor discrepancy becomes apparent only for the coarsest configuration ($\Delta x = D/12$) compared to the finer grids, indicating that solution accuracy begins to degrade at this resolution.  

The computational cost of the model scales approximately linearly with the total number of grid points ($N_x \times N_r$), making it possible to achieve significant efficiency gains through mesh coarsening. Using a 1\% relative velocity increment as the convergence criterion, the computational time reduces from approximately 500 seconds for the finest grid ($\Delta x = D/20$) to merely 140 seconds for the coarsest grid ($\Delta x = D/12$). When employing a more stringent tolerance with normalized residual of $10^{-8}$ as the stopping criterion, the computational requirements approximately triple: the finest grid solution requires 1420 seconds, while the coarsest grid solution completes in 410 seconds. All simulations were performed using a Python implementation of the model with the SciPy sparse linear algebra solvers on a 24-core processor operating at 2.5 GHz.  


\begin{figure}
    \centering

    \includegraphics[width=0.75\linewidth]{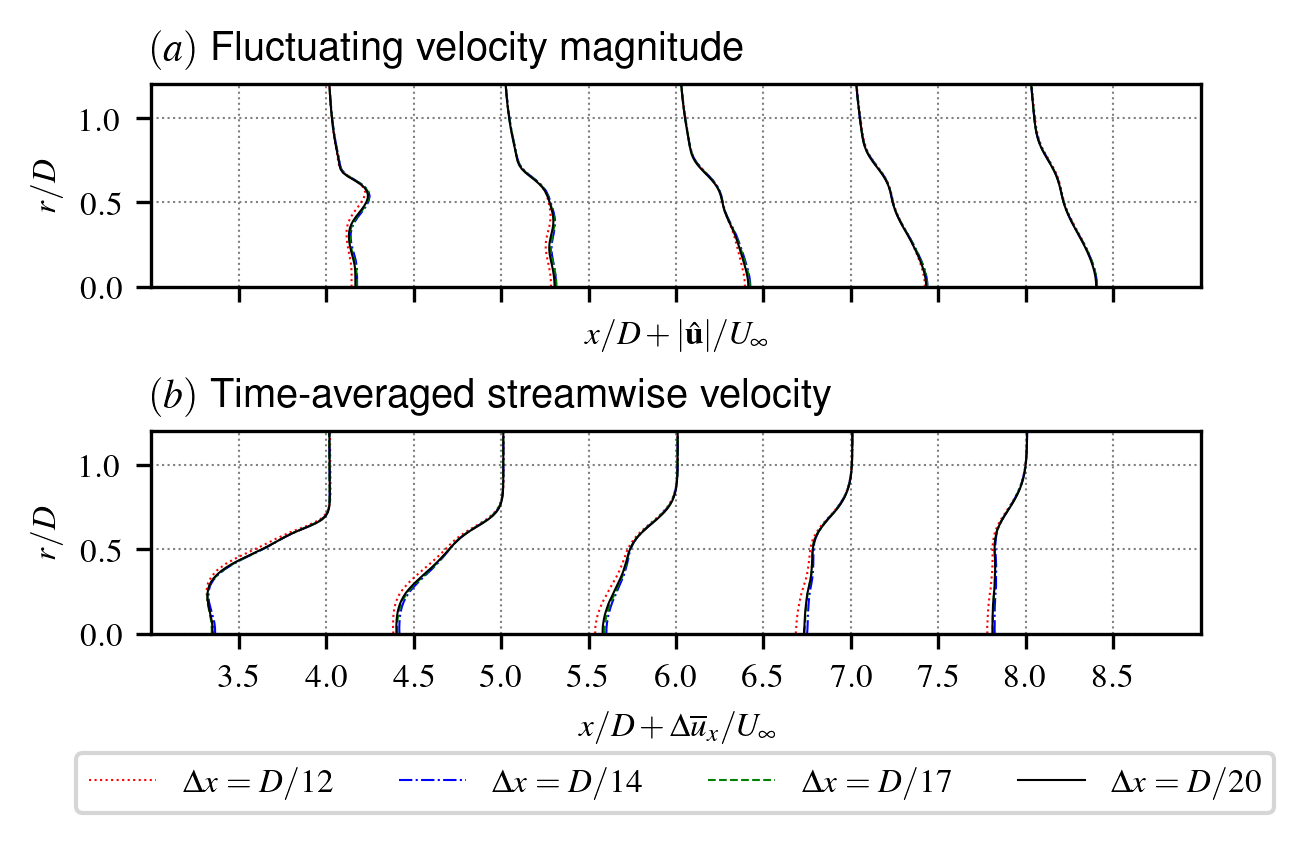}
    \caption{Effect of grid resolution on self-consistent model solutions : (a) total fluctuating velocity magnitude, (b) time-averaged streamwise velocity. Results are plotted at downstream locations $x/D \in \{4,5,6,7,8\}$. AWC approach: DIPC with co-rotating helix St = 0.25.}
    \label{fig:meshConvergence}
    
\end{figure}

}

\bibliographystyle{jfm}
\bibliography{jfm}

\end{document}